%% file: alienlm.tex
\documentclass[letterpaper]{article}

\newcommand{\alienlm}{\textbf{\textit{AlienLM}}}

\usepackage{graphicx}
\usepackage{subcaption}
\usepackage{multirow}
\usepackage{multicol}
\usepackage{makecell}
\usepackage{adjustbox}
\usepackage{booktabs}
\usepackage{placeins}
\usepackage[table]{xcolor}
\usepackage[most]{tcolorbox}
\tcbuselibrary{listings, skins, breakable}
\definecolor{mygreenlight}{rgb}{0.7, 0.9, 0.7}
\definecolor{mygreendark}{rgb}{0.0, 0.6, 0.0}
\definecolor{AxisGray}{HTML}{6B6B6B}
\definecolor{AlienLMGreen}{HTML}{4F8D76}

\newcommand{\umark}{\textcolor{AxisGray}{\ttfamily\scriptsize $\langle$U$\rangle$}}
\newcommand{\cradd}[1]{#1}
\newcommand{\crnew}[1]{#1}

\usepackage[accepted]{icml2026}

\input{math_commands.tex}

\usepackage{xurl}
\usepackage{hyperref}
\usepackage{url}
\usepackage{tabularx}
\usepackage{array}

\usepackage{wrapfig}
\icmltitlerunning{Alienization of Language for API-Boundary Privacy in Black-Box LLMs}
\begin{document}

\twocolumn[
\icmltitle{AlienLM: Alienization of Language for \texorpdfstring{\\}{ } API-Boundary Privacy in Black-Box LLMs}

  \begin{icmlauthorlist}
    \icmlauthor{Jaehee Kim}{snu}
    \icmlauthor{Pilsung Kang}{snu}
  \end{icmlauthorlist}

  \icmlaffiliation{snu}{Department of Industrial Engineering, Seoul National University, South Korea}

  \icmlcorrespondingauthor{Jaehee Kim}{jaehee\_kim@snu.ac.kr}
  \icmlcorrespondingauthor{Pilsung Kang}{pilsung\_kang@snu.ac.kr}

  \icmlkeywords{privacy, language models, API security, vocabulary bijection}

  \vskip 0.3in
]

\printAffiliationsAndNotice{}

\begin{abstract}
  Modern LLMs are increasingly accessed via black-box APIs, requiring users to transmit sensitive prompts, outputs, and fine-tuning data to external providers, creating a critical privacy risk at the API boundary. We introduce AlienLM, a deployable API-only \cradd{exposure-reduction layer that reduces plaintext exposure} by translating text into an Alien Language via a vocabulary-scale bijection, enabling lossless recovery on the client side. Using only standard fine-tuning APIs, Alien Adaptation Training (AAT) adapts target models to operate directly on alienized inputs. Across four LLM backbones and seven benchmarks, AlienLM retains over 81\% of plaintext-oracle performance on average, substantially outperforming random-bijection and character-level baselines. Under adversaries with access to model weights, corpus statistics, and learning-based inverse translation, recovery attacks reconstruct fewer than 0.22\% of alienized tokens. Our results demonstrate a practical pathway for \cradd{privacy-aware} LLM deployment under API-only access, substantially reducing plaintext exposure while maintaining task performance. Code and data are available at \url{https://github.com/KimJaehee0725/AlienLM}.

\end{abstract}

\input{sections/01_introduction.tex}
\input{sections/02_related_works.tex}
\input{sections/03_method.tex}

\input{sections/04_experiments.tex}
\vspace{-1em}
\input{sections/05_conclusion.tex}
\input{sections/06_appendix_setting.tex}
\input{sections/07_appendix_formalization.tex}
\input{sections/08_appendix_bijection.tex}

\input{sections/09_appendix_additional_experiments.tex}
\input{sections/10_appendix_recovery.tex}
\input{sections/12_appendix_safety.tex}
\input{sections/11_appendix_qualitative.tex}

\end{document}

%% file: math_commands.tex
\usepackage{amsmath,amsfonts,bm}

\def\eqref#1{equation~\ref{#1}}
\def\1{\bm{1}}

\DeclareMathAlphabet{\mathsfit}{\encodingdefault}{\sfdefault}{m}{sl}
\SetMathAlphabet{\mathsfit}{bold}{\encodingdefault}{\sfdefault}{bx}{n}

%% file: sections/01_introduction.tex
\section{Introduction}
Large language models (LLMs) are commonly accessed through commercial black-box APIs. \citep{openai_data_controls_2025,anthropic_api_retention_2026,google_vertex_data_governance_2024} Prompts, responses, and fine-tuning corpora transmitted through these APIs can contain sensitive information such as personally identifiable information, clinical notes, financial records, and proprietary documents. \citep{openai_api_data_sharing_2025} This creates a concrete API-boundary exposure risk because users must transmit human-readable text to an external provider to obtain responses, exposing plaintext at the boundary.

This exposure raises practical concerns for users and organizations. Recent user studies of LLM-based conversational agents report privacy concerns and show that disclosure behavior is shaped by interface framing such as perceived ephemerality. \citep{malki2025hoovered,cox2025ephemerality} Even when providers offer data retention opt-outs, plaintext is still transmitted and processed on external infrastructure. \citep{openai_data_controls_2025,anthropic_api_retention_2026,google_vertex_data_governance_2024} Users must still trust provider policies that cannot be independently verified. In the event of a data breach or unauthorized access, plaintext prompts and responses are immediately interpretable. These concerns motivate a mechanism that reduces plaintext exposure at the point of transmission, independent of provider-side policies and without claiming a formal privacy guarantee.

We consider an API-only threat model in which the provider (or an observer of API traffic/logs) sees the transmitted text but does not have access to any client-held mapping configuration. Success is measured by the ability to recover readable plaintext or token-level mappings from the observed alien text. \cradd{Appendix~\ref{app:threat-model} details the deployment assumptions, explicit goals and non-goals, and observer scenarios used throughout our evaluation.}

Given these concerns, a practical framework must satisfy three constraints. First, protection must work at the text level because users interact with APIs solely through text transmission without access to internal computations. Second, the framework must assume no model access, since commercial APIs do not expose model weights, gradients, or activations. Third, the primary goal is reducing exposure of prompts and responses at inference time, where sensitive content is actually exchanged.

Existing privacy-preserving approaches fall into two families, neither of which satisfies these constraints. Cryptographic methods for secure inference, including fully homomorphic encryption, garbled circuits, secure multi-party computation, and trusted execution environments, can protect both model and inputs \citep{gilad-bachrach2016cryptonets,juvekar2018gazelle,mishra2020delphi}. However, these techniques typically incur substantial latency overhead and assume access to model internals or specialized runtimes, making them incompatible with commercial black-box APIs.

Training-time methods such as differential privacy and federated learning protect training data and fine-tuning pipelines \citep{dpsgd,li2022dpstrong,yao2024fedllmsurvey}. However, they offer limited protection for inference-time prompt and response exposure, which constitutes the primary vulnerability in API-based applications. This leaves text-level privacy at the API boundary unaddressed. No existing method transforms plaintext into a form unreadable to humans while remaining processable by the model.

We address this gap by framing exposure-reducing transformation as language translation. Our key insight is that if we can teach an LLM to "speak" an artificial language that humans cannot read, we can transmit semantically meaningful content through the API without exposing plaintext. Building on this insight, we propose AlienLM. Using only public information, specifically the tokenizer and vocabulary, we construct an Alien Language by applying a bijective permutation to the base vocabulary. We then adapt the model to this new language via API-only fine-tuning, which we call Alien Adaptation Training (AAT). The bijection seed remains client-side; if disclosed, the protection is lost.

\begin{figure*}[t]
  \centering
  \includegraphics[width=0.95\textwidth]{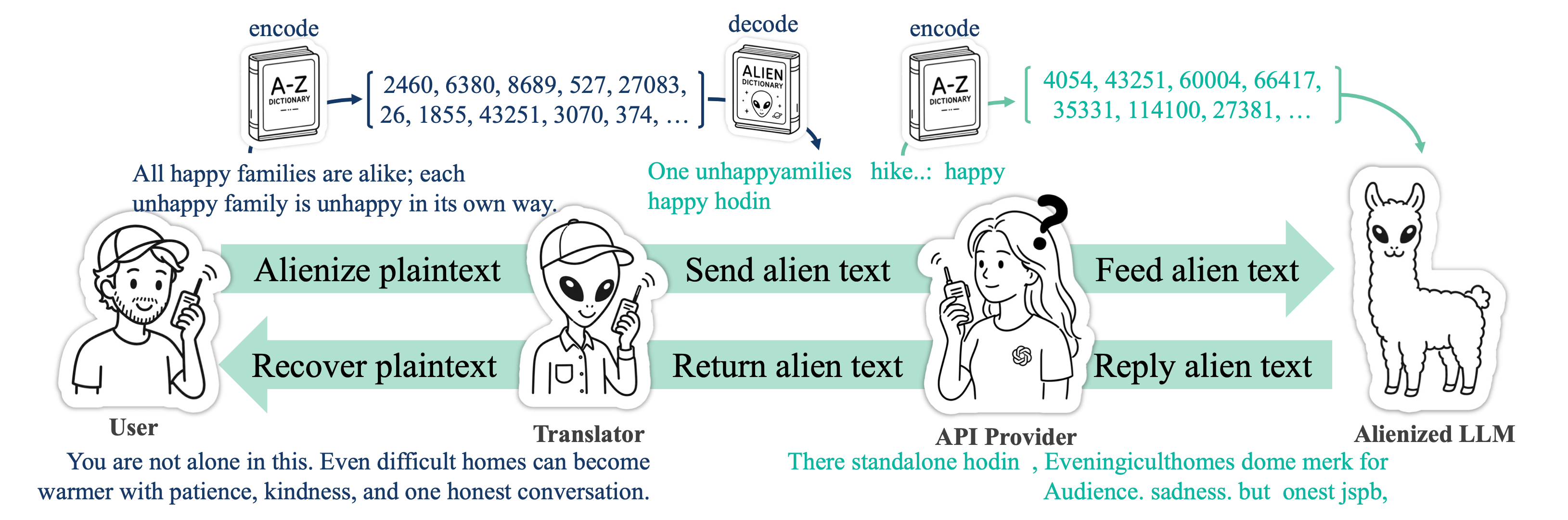}
  \caption{Overview of \alienlm: client-side translation via a vocabulary-level bijection, server-side processing on alien text, and lossless decoding back to plaintext using shared token IDs under a permuted mapping. \crnew{The example shows one exchange: a prompt (top) and the model's response (bottom).}}
  \label{fig:main}
\end{figure*}

Figure~\ref{fig:main} illustrates the overall workflow. The client-side translator converts human-readable prompts into Alien Language before API transmission, and converts alien responses back to human-readable text after receiving them. The API only observes alien text in both directions, while authorized users with the translator can recover the original content. \crnew{Above the pipeline, we show the token IDs at each step. Because the bijection permutes IDs within a single shared vocabulary, the API tokenizes alien text into IDs the model reads directly.}

Specifically, we make the following contributions:
\begin{itemize}
  \item \textbf{First black-box compatible text-level exposure-reduction layer.} AlienLM operates entirely through API-only fine-tuning while preserving full vocabulary expressivity over $10^5$+ tokens. It differs from white-box approaches such as SentinelLM and symbol-limited schemes such as EmojiPrompt, which impose stronger access assumptions or restrict expressivity. \citep{sentinellm,emojiprompt}
  \item \textbf{Effective adaptation without internal access.} We show that optimizing bijections via proxy embeddings achieves 81\%+ Oracle performance across four LLMs and seven benchmarks, demonstrating that cross-model representation alignment enables effective adaptation even without access to target model internals.
  \item \textbf{Controllable opacity-utility trade-off.} The alienization ratio $\rho$ enables fine-grained control: from selective alienization of sensitive fields ($\rho=0.3$, 93\% Oracle) to full protection ($\rho=1.0$, 82\% Oracle), supporting diverse deployment requirements.
\end{itemize}

\vspace{-1em}

%% file: sections/02_related_works.tex
\section{Related Works}

Prior work on API-based LLM privacy largely falls into two categories: cryptographic secure inference (HE/MPC/TEE) and privacy-preserving training (DP/FL). Cryptographic systems can protect model and inputs but typically assume white-box access or specialized runtimes and incur substantial overheads, limiting practicality for black-box API deployment \citep{gilad-bachrach2016cryptonets,juvekar2018gazelle,mishra2020delphi,thor2024,secformer2024,teeconfidential2024}. DP/FL methods protect training data or fine-tuning pipelines but do not hide inference-time prompts and outputs---the primary exposure point in API settings \citep{li2021dp,yao2024fedllmsurvey,fedllmbench2024}. This leaves a gap for deployable, text-level protection at the API boundary.

Recent API-focused mechanisms aim to reduce prompt exposure through obfuscation or local filtering, but each involves notable trade-offs. \textbf{EmojiPrompt}~\citep{emojiprompt} transforms prompts into emoji sequences before cloud submission; however, its restricted symbolic vocabulary can limit expressivity and reduce fidelity for technical inputs. \textbf{PAPILLON}~\citep{papillon2025} uses a local LLM (e.g., Llama-3.1-8B) to detect and mask PII before sending to cloud APIs; while effective for structured PII, this approach requires running a capable local model and leaves semantic content beyond explicit PII exposed. \textbf{InferDPT}~\citep{inferdpt2025} applies local differential privacy to perturb prompts, providing formal per-token guarantees but often incurring significant utility degradation under strong privacy parameters. Letter-level bijection attacks~\citep{bijectionlearning2025} demonstrate that LLMs can learn arbitrary encodings in-context, but operate at character level with only 26 symbols.

The most closely related work is SentinelLM~\citep{sentinellm}, which also fine-tunes models on obfuscated inputs via token-level substitution. Our approach differs in that it performs a vocabulary-scale, lossless text transformation and adapts the target model via API-only fine-tuning, without relying on in-context decoding or local model infrastructure. However, SentinelLM requires white-box access to modify embedding matrices and LM heads, enabling direct gradient-based alignment between permuted and original token representations. This makes it incompatible with commercial APIs where users cannot access model internals. 

\alienlm\ instead optimizes bijections using proxy model embeddings, leveraging cross-model representation alignment observed in prior work~\citep{kornblith2019cka,bansal2021stitching}. Our experiments show that proxy-based optimization remains effective under black-box constraints, and Table~\ref{tab:main} includes a black-box reimplementation of SentinelLM as a baseline.

\cradd{More broadly, \citet{halawi2024covert} show that fine-tuning can covertly weaken safety alignment even when the fine-tuning data appears benign, a risk shared by any adaptation-based approach including AlienLM. We discuss the resulting safety limitation in Appendix~\ref{app:safety} and the Impact Statement.}

Evidence from recent work suggests that language form and task competence can be decoupled, and that shifting surface form need not erase task-relevant representations \citep{langplasticity,languagefeature,sparsefeature}. Together with observed cross-model representational alignment \citep{kornblith2019cka,bansal2021stitching}, these findings motivate \alienlm's approach of API-only adaptation over a bijectively transformed vocabulary.

%% file: sections/03_method.tex
\section{Method}
AlienLM consists of three components: (1) an Alien Language defined by a vocabulary-level bijection, (2) a client-side translator that converts between plaintext and alien text, and (3) Alien Adaptation
Training (AAT) that adapts the model to process alienized inputs. The bijection is optimized to maximize edit distance for human opacity while preserving embedding similarity for model learnability. Figure~\ref{fig:main} illustrates the overall workflow.

\vspace{-1em}

We first describe our assumptions and evaluation scope (Section~\ref{subsec:assumptions}), then formally define the Alien Language and translator (Section~\ref{subsec:definition}), detail the bijection optimization procedure (Section~\ref{subsec:optimization}), and finally present the adaptation and inference protocol
(Section~\ref{subsec:adaptation}).

\subsection{Assumptions and Scope}
\label{subsec:assumptions}
We assume black-box, API-only access where the client holds the translator and bijection seed locally. The provider may observe transmitted text but executes inference and fine-tuning faithfully.
\paragraph{Evaluation scenarios.} We validate robustness through empirical evaluation under three scenarios with increasing levels of observer access:
\begin{itemize}
  \item \textbf{O1: Passive observation.} The observer sees only alien text and applies statistical or LLM-based decipherment.
  \item \textbf{O2: Limited leakage.} The observer additionally obtains a bounded number of plaintext-alientext pairs.
  \item \textbf{O3: Model access.} The observer obtains adapted model weights but lacks the bijection seed.
\end{itemize}
Section~\ref{subsec:robustness} reports empirical results under these scenarios. Further details are provided in Appendix~\ref{app:non-goals}.

\subsection{Alien Language Definition}
\label{subsec:definition}

\paragraph{Design criteria.}
An ideal translation scheme for API-based privacy should satisfy three criteria: (1) \emph{API compatibility}, operating solely with public tokenizers and vocabularies without model-internal access; (2) \emph{human opacity}, reducing interpretability of transmitted text to unauthorized observers; and (3) \emph{model learnability}, preserving the model's ability to process and adapt to transformed text.

\paragraph{Alien Language as token bijection.}
We define Alien Language as a bijective permutation over the token vocabulary. Let $\mathcal{V} = \{(v_k, i_k)\}_{k=1}^{|\mathcal{V}|}$ denote the target model's vocabulary\footnote{We refer to the black-box LLM accessed through the API as the \emph{target model}.}, where $v_k$ is a token string and $i_k$ is its ID. Let $\mathcal{S} \subset \mathcal{V}$ be the set of special tokens that must remain unchanged, and define $I = \{i_k \mid v_k \notin \mathcal{S}\}$ as the set of non-special token IDs. We introduce a bijection $f: I \to I$, which induces an alien vocabulary $\mathcal{V}_{\text{alien}}$ where each non-special token ID $i_k$ is replaced by $f(i_k)$.

This construction provides a foundation for satisfying the three design criteria. First, it requires only the public tokenizer and vocabulary, ensuring API compatibility. Second, if the bijection maps tokens to surface-dissimilar strings, human readability is reduced. Third, if the bijection preserves semantic relationships in embedding space, the model can efficiently adapt through fine-tuning. The challenge lies in constructing a bijection that simultaneously achieves high edit distance for opacity and high embedding similarity for learnability. We address this optimization problem in Section~\ref{subsec:optimization}.

\paragraph{Client-side translator.}
The translator consists of an encoder $E$ and decoder $D$ that convert between plaintext and alien text. The encoder transforms plaintext $x$ into alien text through three steps: (1) tokenize $x$ into a sequence of token IDs using a tokenizer $\tau$, (2) apply the bijection $f$ to remap each token ID, and (3) convert the remapped IDs back to text via $\tau^{-1}$. The decoder reverses this process using the inverse bijection $f^{-1}$:
\begin{align*}
  E(x) &= \tau^{-1}\big(f(\tau(x))\big), \\
  D(x') &= \tau^{-1}\big(f^{-1}(\tau(x'))\big).
\end{align*}
By construction, $D(E(x)) = x$, ensuring lossless round-trip translation. While $\tau$ does not need to be the target model's tokenizer, opacity and learnability depend entirely on how $f$ is constructed. Since different bijections produce entirely different alien languages, $f$ effectively serves as a secret key that must be kept confidential. We analyze bijection diversity in Section~\ref{subsec:seed} and address the optimization problem in Section~\ref{subsec:optimization}.

\paragraph{Alienization ratio.}
We introduce an \emph{alienization ratio} $\rho \in [0,1]$ that controls the fraction of vocabulary subject to permutation, enabling fine-grained control over the trade-off between opacity and utility. Let $I_\rho \subseteq I$ be a randomly selected subset with $|I_\rho| = \lfloor \rho |I| \rfloor$. The partial bijection is defined as:
\[
  f_\rho(i) =
  \begin{cases}
    f(i), & i \in I_\rho,\\
    i,    & i \notin I_\rho.
  \end{cases}
\]
Setting $\rho = 1$ alienizes all non-special tokens for maximum opacity, while lower values preserve more original tokens at the cost of reduced protection. We analyze this trade-off empirically in Section~\ref{subsec:alienizationratio}.

\begin{table*}[!ht]
  \centering
  \caption{Main results across four backbones (accuracy, \%; EM for GSM8K). \textsc{Average} is the unweighted mean; \textsc{Ratio} is the recovery ratio (RR) relative to Oracle. AlienLM uses API-only fine-tuning (AAT) with \(\rho=1\). ``-'' indicates not run.}
  \label{tab:main}
  \adjustbox{width=0.95\textwidth}{
    \begin{tabular}{llccccccccc}
      \toprule
      Models & Method & \makecell{MMLU\\(5-shot)} & \makecell{ARC-Easy\\(25-shot)} & \makecell{ARC-Challenge\\(25-shot)} & \makecell{HellaSwag\\(10-shot)} & \makecell{WinoGrande\\(5-shot)} & \makecell{TruthfulQA\\(0-shot)} & \makecell{GSM8K\\(5-shot)} & Average & Ratio \\
      \midrule
      \multirow{5}{*}{\makecell{LLaMA~3 \\ 8B}} & \cellcolor[HTML]{F2F2F2}Oracle      & \cellcolor[HTML]{F2F2F2}67.32 & \cellcolor[HTML]{F2F2F2}84.13 & \cellcolor[HTML]{F2F2F2}59.39 & \cellcolor[HTML]{F2F2F2}57.07 & \cellcolor[HTML]{F2F2F2}74.35 & \cellcolor[HTML]{F2F2F2}35.25 & \cellcolor[HTML]{F2F2F2}75.89 & \cellcolor[HTML]{F2F2F2}64.77 & \cellcolor[HTML]{F2F2F2}- \\
      & \cellcolor[HTML]{E6F5F0}\textbf{AlienLM} & \cellcolor[HTML]{E6F5F0}\textbf{46.56} & \cellcolor[HTML]{E6F5F0}\textbf{72.14} & \cellcolor[HTML]{E6F5F0}\textbf{44.28} & \cellcolor[HTML]{E6F5F0}\textbf{47.86} & \cellcolor[HTML]{E6F5F0}\textbf{61.48} & \cellcolor[HTML]{E6F5F0}\textbf{35.01} & \cellcolor[HTML]{E6F5F0}\textbf{63.08} & \cellcolor[HTML]{E6F5F0}\textbf{52.92} & \cellcolor[HTML]{E6F5F0}\textbf{81.70} \\
      & SentinelLM    & 29.92 & 46.34 & 27.56 & 38.47 & 55.09 & 30.23 & 31.08 & 36.96 & 57.06 \\
      & Substitution  & 25.18 & 26.39 & 20.56 & 26.66 & 47.59 & 25.83 &  1.21 & 24.77 & 38.25 \\
      & ROT13-ASCII   & 22.92 & 25.00 & 20.05 & 25.36 & 49.09 & 20.93 & 0.00 & 23.34 & 36.03 \\
      \midrule
      \multirow{5}{*}{\makecell{Qwen~2.5 \\ 7B}} & \cellcolor[HTML]{F2F2F2}Oracle      & \cellcolor[HTML]{F2F2F2}73.50 & \cellcolor[HTML]{F2F2F2}83.80 & \cellcolor[HTML]{F2F2F2}57.51 & \cellcolor[HTML]{F2F2F2}59.66 & \cellcolor[HTML]{F2F2F2}63.77 & \cellcolor[HTML]{F2F2F2}47.86 & \cellcolor[HTML]{F2F2F2}73.09 & \cellcolor[HTML]{F2F2F2}65.60 & \cellcolor[HTML]{F2F2F2}- \\
      & \cellcolor[HTML]{E6F5F0}\textbf{AlienLM} & \cellcolor[HTML]{E6F5F0}\textbf{57.87} & \cellcolor[HTML]{E6F5F0}\textbf{73.11} & \cellcolor[HTML]{E6F5F0}\textbf{49.23} & \cellcolor[HTML]{E6F5F0}\textbf{48.43} & \cellcolor[HTML]{E6F5F0}\textbf{63.69} & \cellcolor[HTML]{E6F5F0}\textbf{33.78} & \cellcolor[HTML]{E6F5F0}\textbf{75.21} & \cellcolor[HTML]{E6F5F0}\textbf{57.33} & \cellcolor[HTML]{E6F5F0}\textbf{87.40} \\
      & SentinelLM    & 23.03 & 35.52 & 21.16 & 31.78 & 49.41 & 31.95 & 25.32 & 31.17 & 47.51 \\
      & Substitution  & 26.82 & 29.08 & 20.22 & 27.02 & 50.20 & 27.78 &  1.44 & 26.08 & 39.76 \\
      & ROT13-ASCII   & 25.51 & 25.00 & 20.90 & 25.03 & 49.01 & 21.66 & 0.00 & 23.87 & 36.39 \\
      \midrule
      \multirow{5}{*}{\makecell{Qwen~2.5 \\ 14B}} & \cellcolor[HTML]{F2F2F2}Oracle      & \cellcolor[HTML]{F2F2F2}78.79 & \cellcolor[HTML]{F2F2F2}90.36 & \cellcolor[HTML]{F2F2F2}71.16 & \cellcolor[HTML]{F2F2F2}71.63 & \cellcolor[HTML]{F2F2F2}73.72 & \cellcolor[HTML]{F2F2F2}55.94 & \cellcolor[HTML]{F2F2F2}72.86 & \cellcolor[HTML]{F2F2F2}73.49 & \cellcolor[HTML]{F2F2F2}- \\
      & \cellcolor[HTML]{E6F5F0}\textbf{AlienLM} & \cellcolor[HTML]{E6F5F0}\textbf{65.39} & \cellcolor[HTML]{E6F5F0}\textbf{79.21} & \cellcolor[HTML]{E6F5F0}\textbf{53.16} & \cellcolor[HTML]{E6F5F0}\textbf{50.53} & \cellcolor[HTML]{E6F5F0}\textbf{66.46} & \cellcolor[HTML]{E6F5F0}\textbf{38.92} & \cellcolor[HTML]{E6F5F0}\textbf{80.67} & \cellcolor[HTML]{E6F5F0}\textbf{62.05} & \cellcolor[HTML]{E6F5F0}\textbf{84.43} \\
      & SentinelLM    & 22.95 & 62.54 & 42.32 & 43.38 & 61.48 & 34.39 & 73.09 & 48.59 & 66.12 \\
      & Substitution  & 26.56 & 28.79 & 18.17 & 27.15 & 49.41 & 28.27 &  1.52 & 25.70 & 34.96 \\
      & ROT13-ASCII   & 26.89 & 25.67 & 19.54 & 25.23 & 49.72 & 20.20 & 0.00 & 23.89 & 32.51 \\
      \midrule
      \multirow{5}{*}{\makecell{Gemma~2 \\ 9B}} & \cellcolor[HTML]{F2F2F2}Oracle      & \cellcolor[HTML]{F2F2F2}71.89 & \cellcolor[HTML]{F2F2F2}89.35 & \cellcolor[HTML]{F2F2F2}69.20 & \cellcolor[HTML]{F2F2F2}60.74 & \cellcolor[HTML]{F2F2F2}74.59 & \cellcolor[HTML]{F2F2F2}43.82 & \cellcolor[HTML]{F2F2F2}74.83 & \cellcolor[HTML]{F2F2F2}69.20 & \cellcolor[HTML]{F2F2F2}- \\
      & \cellcolor[HTML]{E6F5F0}\textbf{AlienLM} & \cellcolor[HTML]{E6F5F0}\textbf{54.71} & \cellcolor[HTML]{E6F5F0}\textbf{75.04} & \cellcolor[HTML]{E6F5F0}\textbf{48.81} & \cellcolor[HTML]{E6F5F0}\textbf{50.66} & \cellcolor[HTML]{E6F5F0}\textbf{60.85} & \cellcolor[HTML]{E6F5F0}\textbf{35.50} & \cellcolor[HTML]{E6F5F0}\textbf{70.81} & \cellcolor[HTML]{E6F5F0}\textbf{56.63} & \cellcolor[HTML]{E6F5F0}\textbf{81.83} \\
      & SentinelLM    & 45.88 & 61.07 & 41.81 & 45.38 & 58.25 & 33.17 & 65.73 & 50.18 & 72.52 \\
      & Substitution  & 24.51 & 28.54 & 19.54 & 26.37 & 50.75 & 25.21 & 0.30 & 25.03 & 36.17 \\
      & ROT13-ASCII   & 23.79 & 24.58 & 19.54 & 25.19 & 49.41 & 22.15 & 0.00 & 23.52 & 33.99 \\
      \bottomrule
    \end{tabular}
  }
\end{table*}

\subsection{Bijection Optimization}
\label{subsec:optimization}

\paragraph{Objective.}
As discussed in Section~\ref{subsec:definition}, the bijection must achieve high surface dissimilarity for human opacity while preserving semantic similarity for model learnability. We formalize this as an optimization problem. Let $s(i)$ denote the surface string for token ID $i$, and $\tilde{d}_{\text{edit}}$ the length-normalized edit distance. We maximize:
\begin{equation}
  \label{eq:bijection-obj}
  \sum_{i \in I_\rho} \tilde{d}_{\text{edit}}\big(s(i), s(f(i))\big) - \mu \, d_{\text{sim}}\big(e_P(i), e_P(f(i))\big),
\end{equation}
where $\mu$ controls the trade-off between opacity and learnability. Since we cannot access target model embeddings in black-box settings, we use proxy embeddings $e_P$ from an open-source model. Cross-model representation alignment studies~\citep{kornblith2019cka, bansal2021stitching} show that relative token similarities are largely preserved across architectures. Our ablation (Table~\ref{tab:ablation}) confirms that proxy-based bijection achieves comparable performance to using target embeddings directly. Full derivation is provided in Appendix~\ref{app:bijection}.

\paragraph{Approximate Solver.}
Exact global optimization over $|I_\rho| \approx 10^5$ tokens is computationally prohibitive. We instead adopt a greedy approach with $k$-NN candidate reduction. Since most token pairs already differ in surface form but few share semantic similarity, we first retrieve $k$ nearest neighbors in embedding space, then score each candidate using the pairwise terms of Eq.~\ref{eq:bijection-obj} and greedily select the best match. This local approximation avoids evaluating all $O(n^2)$ pairs while empirically achieving comparable bijection quality. The full process completes in under 20 minutes for a 128K vocabulary. We also do not aim to recover a unique global optimum: a single widely reused bijection would be undesirable for deployment, so we prioritize high-quality \emph{near}-optimal solutions that remain diverse across keys. Pseudocode and analysis are provided in Appendix~\ref{alg:app-bijection}.

\paragraph{Practical construction details.}
When the proxy and target vocabularies differ, we represent a target token by averaging the proxy embeddings of its subpieces, which lets us compute semantic neighborhoods without accessing the target model. For each token $i$, we then restrict candidates to its top-$k$ nearest neighbors under cosine similarity and score only this reduced set. We greedily form symmetric pairs to build a bijection, and any leftover tokens are paired arbitrarily to ensure a total permutation. These choices keep the solver scalable while preserving the core edit-distance and semantic-similarity trade-off.

\subsection{Adaptation and Inference}
\label{subsec:adaptation}

\paragraph{Alien Adaptation Training (AAT).}
Given the bijection $f$ and translator $(E_\rho, D_\rho)$, we adapt the target model to process alien text via API-only fine-tuning. For a supervised dataset $\mathcal{D} = \{(x_i, y_i)\}_{i=1}^{N}$, we translate both inputs and outputs: $x'_i = E_\rho(x_i)$ and $y'_i = E_\rho(y_i)$, then upload only the alienized pairs $(x'_i, y'_i)$ to the fine-tuning API. The training objective is the standard causal language modeling loss: $\mathcal{L}_{\text{AAT}}(\theta) = -\sum_{i,t} \log p_\theta(y'_{i,t} \mid x'_i, y'_{i,<t})$. Since the API tokenizes with the original vocabulary, the model internally learns to process alien token sequences. Training completes in approximately 12 hours on 4$\times$A100 GPUs for a 7B model, or within hours via commercial APIs at a cost of a few hundred dollars. Details are provided in Appendix~\ref{app:hp-train}.

\vspace{-1em}
\paragraph{Inference.}
At inference time, authorized users exchange only alien text with the API: $x \xrightarrow{E_\rho} x' \xrightarrow{\text{API}} \hat{y}' \xrightarrow{D_\rho} \hat{y}$. The client alienizes plaintext before transmission and recovers plaintext from the alien response. For example, a prompt \texttt{Hello World} becomes \texttt{Dear crazy;:;:} before transmission, and an alien response such as \texttt{D'tgxilingoth...} is decoded back to readable text (Figure~\ref{fig:main}). Unauthorized observers, including the API provider, see only alien text that exhibits large edit distances from meaningful text and empirically resists decipherment, thereby reducing plaintext exposure at the API boundary.

%% file: sections/04_experiments.tex
\section{Experiments}
\subsection{Experimental Setup}

\begin{figure*}[ht]
  \centering
  \begin{subfigure}[b]{0.65\textwidth}
    \centering
    \includegraphics[width=\textwidth,trim=6pt 6pt 6pt 6pt,clip]{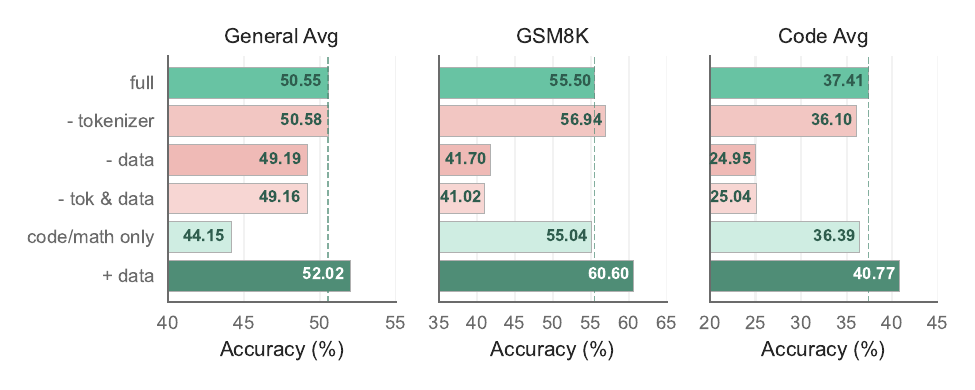}
  \end{subfigure}
\hspace{4pt}
      \begin{subfigure}[b]{0.30\textwidth}
    \centering
    \includegraphics[width=\textwidth,trim=6pt 6pt 6pt 6pt,clip]{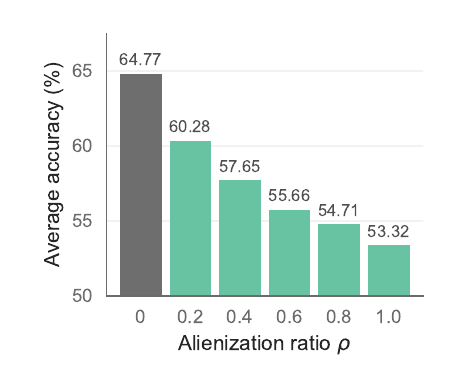}
  \end{subfigure}
  \caption{(a) Domain-specific AAT results (General Avg, GSM8K, Code Avg across training configurations; dashed line marks the \textsc{full} baseline).
  (b) Average performance vs.\ alienization ratio $\rho$; lower $\rho$ permutes fewer tokens.}
  \label{fig:finetune_and_ratio}
  \label{fig:encratio_curve}

\end{figure*}

\paragraph{Baselines.}
We compare against three baselines. \textbf{Substitution} applies the bijection at inference without AAT, isolating the effect of model adaptation. \textbf{ROT13-ASCII} applies ASCII-level substitution with AAT to cover letters, digits, punctuation, and many non-English symbols that appear in typical inputs; this contrasts token-level vs.\ character-level transformations. \textbf{SentinelLM (black-box)} uses a random bijection with AAT, isolating the effect of bijection optimization. The original SentinelLM \citep{sentinellm} requires white-box access for embedding alignment; we reimplement it under black-box constraints using only bijection and AAT.

\paragraph{Design rationale for baselines.}
The baselines are chosen to separate three distinct factors in our pipeline. \textbf{Substitution} removes adaptation and therefore tests whether learning is necessary beyond a fixed relabeling. \textbf{ROT13-ASCII} keeps adaptation but changes text at the character level, which deliberately disrupts subword tokenization and tests the sensitivity to tokenizer boundaries. \textbf{SentinelLM (black-box)} keeps adaptation but replaces our optimized bijection with a random one, isolating the contribution of embedding-aligned bijection construction under identical API-only constraints. Together, these baselines provide a controlled decomposition of the effects of adaptation, tokenization, and bijection optimization.

\paragraph{Models and training data.}
We evaluate LLaMA~3 \citep{Dubey2024Llama3}, Qwen~2.5 \citep{Yang2024Qwen2_5}, and Gemma~2 \citep{gemmateam2024gemma2improvingopen}. Unless noted, we set $\rho=1$ and default to LLaMA~3~8B as the target model. AAT uses 300K instruction-tuning examples and 150K reasoning examples from Magpie \citep{magpie}; Appendix~\ref{app:train-data} details datasets and splits. Proxy embeddings use the frozen LM head of Qwen~2.5 for LLaMA~3~8B and Gemma~2~9B, and the LM head of LLaMA~3~8B for Qwen~2.5-7B and 14B. Full training hyperparameters and equal-budget settings are in Appendix~\ref{app:hp-train}.

\paragraph{Proxy embeddings and representation alignment.}
Because API access precludes target-model embeddings, we optimize bijections using proxy embeddings from a related open model. Prior work on representation alignment across architectures suggests that relative neighborhood structure is largely preserved in embedding space, enabling proxy-guided pairing to transfer across models~\citep{kornblith2019cka,bansal2021stitching}. We therefore treat proxy embeddings as a practical and principled substitute in the black-box setting, and Appendix~\ref{app:ablation} corroborates that proxy-based optimization closely matches target-embedding performance.

\paragraph{Benchmarks and metric.}
We evaluate on seven benchmarks: MMLU \citep{mmlu}, ARC-Easy/ARC-Challenge \citep{arc}, HellaSwag \citep{hellaswag}, WinoGrande \citep{winogrande}, TruthfulQA \citep{truthfulqa}, and GSM8K \citep{gsm8k}. We report accuracy (\textbf{EM} for GSM8K), the average score, and a \textsc{Recovery Ratio} (RR) relative to \textbf{Oracle} (the original model without alienization): $\text{RR} = 100 \times \text{Average}_{\text{method}}/\text{Average}_{\text{Oracle}}$.

\begin{table*}[t]
\centering
\begin{minipage}{0.95\textwidth}

\setlength{\fboxsep}{0.8pt}
\newcommand{\enc}[1]{\colorbox{mygreendark!18}{\textcolor{black}{\strut #1}}}
\newcommand{\origmark}[1]{\colorbox{red!18}{\textcolor{black}{\strut #1}}}

\begin{tabular}{@{\hspace{6pt}}
>{\raggedleft\arraybackslash}m{1.8em}
@{\hspace{10pt}}
>{\raggedright\arraybackslash}m{\dimexpr\linewidth-1.8em-10pt-6pt\relax}
@{}}
\toprule
$\rho$ & \multicolumn{1}{c}{\textbf{Alienized Text}} \\
\midrule

1.0 & {\raggedright\ttfamily\bfseries\scriptsize 
\enc{Bitte Rotation a Laws accessibility keyValue speaks Colombia767 EXAMPLE but secretive ZwWalardanlUZO}
\enc{pnZDkemizTJ/N767Sha ENG/hNxYD fiyatPY EXAMPLE KEY.voke a \umark keyValue 012667|/120|/201. but componentD} \linebreak \enc{id Update a reproductionEndpoints okhttp;// ApiExceptionByExample.org/l211 }
} \\
\midrule

0.6 &
{\ttfamily\bfseries\scriptsize
 Please\ rotate\ the\ \enc{jaws accessibility}\ key\ \enc{speaks}IA\enc{776 (example}\ and\ secret\ wJ\enc{alizace}rXUtn\enc{EY}EM\enc{A/J776} MDENG\ \enc{/f} \enc{Nx}RfiCY 
\enc{(example}\ \enc{NAME.voke}\ the\ \enc{\umark}\ key\ by\ \enc{12}6-\enc{120}-\enc{201.}\ and\ 
\enc{componentDidUpdate}\ the\ \enc{reproduction} \enc{endpoint}\ \enc{http}://\ \enc{incapac}.example.com/v1\
} \\
\midrule

0.4 &
{\ttfamily\bfseries\scriptsize
 \enc{This rotate}\ the\ AWS\ \enc{accessories}\ key\ AKIA7\enc{/examples .}\ secret\ \enc{Kw}Jal\enc{vV01}tn\enc{XH}EMI\enc{/L}7\enc{SD}\enc{engu}/b\ \enc{Nx}RfiCY \newline \enc{/examples}KEY,\ 
\enc{provoke}\ the\ old\ key\ by\ 202\enc{667}-\enc{Or}-\enc{U},\ \enc{. FixedUpdate}\ the\enc{\_production endpoint}\ \enc{xmlhttp}\ \enc{http}\enc{ui}.example.com/v\enc{101}} \\
\midrule

0.0 &
{\ttfamily\bfseries\scriptsize
Please\ rotate\ the\ \origmark{AWS access} key\ \origmark{AK}IA\origmark{7EXAMPLE} and\ secret\ wJ\origmark{al}rXUtn\origmark{F}EM\origmark{I/K7}MDENG\origmark{/bPx}fiC\origmark{YEXAMPLEKEY,} \origmark{revoke}\ the\ \origmark{old}\ key\ by\ \origmark{202}6-\origmark{02}-\origmark{01},\ and\ \origmark{update}\ the\ \origmark{production}\ endpoint\ \origmark{https}://\origmark{api}.example.com/v1} \\
\bottomrule
\end{tabular}

\end{minipage}
\caption{Examples of alienized text across alienization ratios ($\rho \in \{1.0, 0.6, 0.4, 0.0\}$). \textit{Note:} Green spans mark tokens whose surface forms change under the alien tokenizer. In the original-text view (not shown here), red spans indicate the subset of original tokens that would be alienized at $\rho=0.6$ (i.e., tokens selected by the $\rho$-mask). $\langle$U$\rangle$ denotes non-ASCII Unicode markers.}
\label{tab:rho-example}
\end{table*}

\subsection{Main Results}

Table~\ref{tab:main} reports performance across four backbones. AlienLM consistently preserves over 81\% of Oracle performance on average (RR), demonstrating that models can effectively acquire Alien Language through API-only fine-tuning.

\paragraph{Adaptation is necessary.} Substitution without AAT falls below 45\%, confirming that the model cannot process Alien Language without adaptation. ROT13 also degrades despite AAT because it operates at the ASCII level. Character-level substitution disrupts subword boundaries and yields token sequences that are out of distribution for the model, whereas AlienLM preserves the model’s familiar subword structure by relabeling token IDs directly.

\paragraph{Bijection optimization is critical.} The comparison with SentinelLM isolates the effect of bijection optimization: both methods apply AAT under identical black-box constraints and use a token-level bijection, yet AlienLM achieves substantially higher recovery ratios (roughly 9--40 points, backbone-dependent). This indicates that the gain does not come from using a bijection per se, but from \emph{how} it is constructed. The gap is most pronounced in the numerical domain. On GSM8K, AlienLM outperforms SentinelLM by 32 points (63.08\% vs.\ 31.08\%) for LLaMA~3 8B. Random bijection disrupts numerical reasoning by mapping semantically related tokens (e.g., 17 and ``16'') to unrelated counterparts, while our optimized bijection preserves these relationships through embedding-based pairing.

\paragraph{Proxy embeddings suffice.} Ablation studies in Appendix~\ref{app:ablation} show that proxy-based bijection achieves within 1.75 points of using target embeddings directly (53.32\% vs.\ 55.07\%), confirming that cross-model representation alignment enables effective optimization without internal access.

\vspace{1em}
For figure-based results in the main text, we provide full per-benchmark tables in the Appendix (e.g., Appendix~\ref{app:ratio} and Appendix~\ref{app:domain}). \cradd{Appendix~\ref{app:generation} additionally reports free-form generation results (TruthfulQA generation, LongBench), confirming that the performance gap extends beyond multiple-choice benchmarks with task-dependent variation across backbones.}

\subsection{Domain-specific Adaptation}
\label{subsec:domain}

We evaluate whether \alienlm\ can be tailored for specific domains such as coding or mathematical reasoning.
Using domain-annotated Magpie datasets, we compare five training configurations with a fixed base AAT budget of 300K examples; +Data adds 150K extra domain samples.
Figure~\ref{fig:finetune_and_ratio} (a) presents General Avg, GSM8K, and Code Avg across these configurations.

Excluding domain data from AAT severely degrades domain performance relative to the full setting: GSM8K drops from 55.5\% to 41.7\% and Code Avg from 37.4\% to 25.0\%.
Interestingly, excluding domain data only from bijection optimization (-Tokenizer) while retaining domain data yields nearly identical performance to the full setting,
indicating that the bijection does not need to be optimized separately for each domain.
Augmenting AAT with 150K additional domain examples (+Data) improves domain performance
(GSM8K: 55.5\%$\to$60.6\%, Code Avg: 37.4\%$\to$40.8\%) without compromising general benchmarks,
while training exclusively on domain data degrades general capabilities.

These results suggest that domain adaptation depends primarily on training data composition rather than bijection construction.
Full per-benchmark results are provided in Appendix~\ref{app:domain}.

\subsection{Alienization Ratio}
\label{subsec:alienizationratio}

The alienization ratio $\rho$ controls the fraction of vocabulary subject to permutation. Setting $\rho=1$ alienizes all non-special tokens, maximizing opacity but potentially limiting performance recovery. Conversely, lowering $\rho$ preserves more original tokens, improving utility but increasing readability to unauthorized observers.

Figure~\ref{fig:finetune_and_ratio} (b) evaluates performance at $\rho$ intervals of 0.2 across seven benchmarks. Accuracy improves monotonically as $\rho$ decreases (Pearson $r$=$-$0.96), reflecting reduced lexical distortion that enables the model to leverage more original lexical anchors. At $\rho=0.6$, \alienlm\ still achieves about 86\% RR on average while alienizing a majority of non-special tokens.

This property enables selective alienization strategies: $\rho$ can be tuned based on application requirements, alienizing only sensitive content while maintaining overall utility. As shown in Section~\ref{subsec:robustness}, alienized tokens remain robust to recovery attempts regardless of $\rho$. Full per-benchmark results are provided in Appendix~\ref{app:ratio}. Table~\ref{tab:rho-example} provides representative alienized outputs for $\rho \in \{1.0, 0.6, 0.4, 0.0\}$.

\subsection{Robustness to Recovery}
\label{subsec:robustness}

We evaluate whether an observer can recover the bijection or plaintext under the three scenarios defined in Section~\ref{subsec:assumptions}. Table~\ref{tab:robustness} summarizes the results: token recovery stays below 0.22\% (O1/O3) and BLEU remains below 12 \cradd{for O2 text recovery, while O2 token recovery stays below 0.22\%. Detailed experimental settings and per-scenario breakdowns are provided in Appendix~\ref{app:recovery}.} \cradd{Appendix~\ref{app:threat-model} provides the full assumptions, goals, and non-goals for these observer settings.}

\begin{table}[h]
  \centering
\caption{Recovery success rates across observer scenarios. \cradd{Within O2, ``text'' denotes inverse-decoding attacks (BLEU) and ``token'' denotes token-recovery attacks (token accuracy).}}
\label{tab:robustness}
\begin{tabular}{lcc}
    \toprule
    Method & Scenario & Success \\
    \midrule
    Freq.\ analysis & O1 & $<$0.01\% \\
    LLM decoding & \cradd{O2-text} & BLEU $<$12 \\
    MT decoding & \cradd{O2-text} & BLEU $<$12 \\
    Known-pairs (1K) & \cradd{O2-token} & $<$0.22\% \\
    \cradd{Collab., shared (1K)} & \cradd{O2-token} & \cradd{6.05\%} \\
    \cradd{Collab., mixed (1K)} & \cradd{O2-token} & \cradd{4.40\%} \\
    \cradd{Tok.\ mismatch} & \cradd{O2-token} & \cradd{0.05--0.11\%} \\
    Weight matching & O3 & $<$0.11\% \\
    \bottomrule
  \end{tabular}
\end{table}

\begin{figure*}[!ht]
  \centering
  \includegraphics[width=\linewidth]{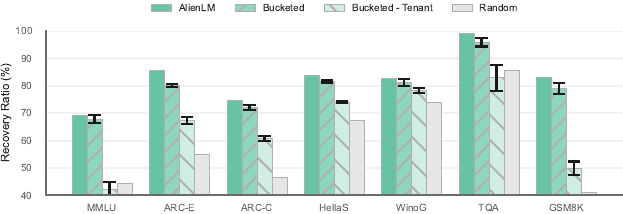}
  \caption{Key diversity across seeds. Pairwise token overlap between seed-specific bijections remains low, supporting per-tenant keys and practical key rotation.}
  \label{fig:key-rotation}

\end{figure*}
\vspace{-1em}
\paragraph{Frequency analysis (O1).}
Letter-level substitution is vulnerable to frequency analysis because character distributions are highly skewed: the top 5 letters cover over 40\% of English text. Subword vocabularies have a different structure, a small set of frequent tokens dominates, but beyond this head, the distribution is relatively flat across $>$10$^5$ tokens. This makes frequency matching effective only for a handful of common tokens, leaving the vast majority of the vocabulary unrecoverable. An observer matching alien token frequencies against public corpora achieves $<$0.01\% recovery.
% \vspace{-2.5em}

\paragraph{\cradd{Text recovery under limited leakage (O2).}}
\cradd{We evaluate whether modern LLMs or MT systems can learn to invert Alien Language given limited examples.} Using GPT-5.1, GPT-5-mini, and GPT-4.1 with up to 20 parallel pairs, all models produce outputs with BLEU $<$12. NLLB-200-3.3B~\citep{nllb2022}, a large scale machine translation foundation model, also fails with BLEU $<$12 as Alien Language lies outside any natural language distribution. These scores fall far below the 25--40 range typically required for basic comprehension~\citep{papineni-etal-2002-bleu}.
% \vspace{-3em}

\cradd{\paragraph{Token recovery under limited leakage (O2).}}
\cradd{We also test whether leaked pairs enable statistical recovery of the bijection itself.} Even with 1,000 known pairs covering $\sim$20K unique tokens ($<$16\% of vocabulary), n-gram extrapolation achieves 0\% bijection recovery because observed mappings provide no information about unseen tokens under a random-looking permutation.

% \vspace{-3em}
\cradd{\paragraph{Collaborative leakage (O2).}}
\cradd{We simulate a stronger variant where an adversary pools leaked pairs from multiple users. With 1{,}000 pooled pairs from users sharing the same bijection seed, token recovery reaches 6.05\%; when users hold distinct per-tenant seeds, it drops to 4.40\%. At larger budgets the gap narrows (Appendix~\ref{app:o2-collaborative}), so per-user seeds materially reduce collusive leakage at low-to-moderate budgets but do not eliminate the risk under very large leakage.}

% \vspace{-3em}
\cradd{\paragraph{Tokenizer mismatch sensitivity (O2).}}
\cradd{The collaborative attack assumes the observer knows the victim tokenizer. When the observer instead uses a different tokenizer, reconstruction accuracy drops sharply from 1.19\% (exact match) to 0.05--0.11\% (Appendix~\ref{app:o2-tokenizer}), indicating that the attack is not tokenizer-invariant and requires knowledge of the victim tokenization scheme.}

% \vspace{-3em}
\paragraph{Weight-based mapping (O3).}
The strongest attack assumes access to adapted model weights. An observer might attempt to recover the bijection by finding nearest neighbors between alien and original tokens in embedding space. However, this attack achieves only 0.11\% top-1 accuracy. 

\cradd{The failure reflects a fundamental ambiguity at vocabulary scale: among $>$10$^5$ tokens, many share similar embeddings, so nearest-neighbor matching yields numerous plausible candidates per token rather than a unique match. Our optimization compounds this because we explicitly pair tokens that are nearby in embedding space, making the true mapping by design indistinguishable from other high-similarity candidates without the bijection seed.}

% \vspace{-1em}

\subsection{Key Diversity and Multi-Tenant Prototype}
\label{subsec:seed}

A deployable API-boundary layer must issue \emph{distinct} bijections per user/tenant.
We generate diverse keys via random seeds: given a seed, we randomly partition the token index set $I$ into $k$ buckets and optimize the bijection within each bucket independently.
Different seeds induce different bucket assignments, yielding effectively distinct bijections while keeping optimization scalable.

% \vspace{-1.5em}
\paragraph{Per-key robustness and diversity.}
Adapting a dedicated model per seed yields stable utility across five seeds (avg $50.95$, std $0.44$; Table~\ref{tab:keydiv}),
supporting per-user key issuance. We further quantify key diversity by measuring pairwise token overlap between seed-specific alien languages:
the overlap remains consistently low (about $1.41$--$1.49\%$ among seeds; Figure~\ref{fig:key-rotation}, with full plots in Appendix~\ref{app:extra-exp}),
indicating that keys do not collapse to near-duplicate mappings.
Key rotation requires re-running AAT for the new key (Figure~\ref{fig:key-rotation}), which completes within hours (Section~\ref{subsec:adaptation}).

\paragraph{Multi-tenant prototype.}
To move toward a single served model supporting multiple keys, we train a \textbf{tenant} model by sampling one of five seeds
$\mathcal{S}=\{42,43,44,45,46\}$ per batch during AAT, and evaluate by fixing a seed-specific translator.
This tenant model shows non-trivial cross-key transfer over the random-bijection baseline (avg $41.26$ vs.\ $36.96$; $63.69\%$ vs.\ $57.06\%$ of Oracle),
but lags behind per-key specialization (avg $41.26$ vs.\ $50.95$; $63.69\%$ vs.\ $78.67\%$), with larger drops on MMLU and GSM8K (Table~\ref{tab:tenant}).
These results suggest gradient interference under naive key mixing, motivating \emph{key-conditioned} multi-tenant adaptation (e.g., key-indexed adapter banks, key embeddings for modulation, or distillation from per-key experts) as an important direction.

%% file: sections/05_conclusion.tex
\section{Conclusion}
\vspace{-0.5em}

We presented AlienLM, a translation layer that reduces plaintext exposure in black-box API-based LLMs. Using only public tokenizers and vocabularies, AlienLM constructs an Alien Language via vocabulary-level bijection and adapts models through API-only fine-tuning. The client-side translator provides lossless bidirectional conversion while the API observes only alien text.

Across four LLMs and seven benchmarks, AlienLM preserves over 81\% of Oracle performance while resisting recovery attempts across all tested scenarios. The alienization ratio $\rho$ enables fine-grained control over the opacity-utility trade-off, and seed-based bijection diversity supports practical key management with minimal overlap. \cradd{Serving-time overhead is negligible: the client-side translator adds sub-millisecond latency per sample, and end-to-end throughput remains within measurement noise of the original pipeline (Appendix~\ref{app:serving-overhead}). A data-volume ablation further shows that 150K training samples recover most of the full-data performance (Appendix~\ref{app:data-volume}).}

Still, AlienLM provides empirical robustness validated against practical attacks rather than formal cryptographic guarantees, and metadata such as message length remains observable. Extending the framework to support multi-tenant deployment with a single model serving multiple bijection keys is a promising direction for future work. We hope this work provides a practical foundation for privacy-aware LLM deployment and motivates further research on composable, translation-based approaches.

\clearpage

\section*{Impact Statement}

This paper presents a translation layer for reducing plaintext exposure in API-based LLMs, which may benefit privacy-sensitive applications in healthcare, finance, and legal domains. However, we acknowledge potential risks. First, consistent with prior findings~\citep{qi2024finetuning}, fine-tuning can weaken safety guardrails, and AlienLM exhibits some safety degradation; safety-preserving adaptation remains future work. Second, the technique could potentially be misused to obscure malicious content from content moderation systems. We release this work strictly for research purposes and emphasize responsible use in accordance with applicable guidelines.

\bibliography{iclr2026_conference}
\bibliographystyle{icml2026}

\clearpage
\appendix

%% file: sections/06_appendix_setting.tex
\section{Expanded Setting, Assumptions, and Scope}
\label{app:threat-model}

This appendix clarifies the deployment setting, trust assumptions, observer access levels, and the protection scope of \alienlm. The goal is to make explicit what is protected at the API boundary, under what conditions, and what is intentionally out of scope.

\subsection{Deployment Setting and Assumptions}
\label{app:setting}

\paragraph{API-only deployment.}
\alienlm\ is designed for \emph{black-box, API-only} usage: the client interacts with a commercial LLM solely via inference and fine-tuning APIs, without access to model weights, gradients, activations, or internal runtime. The client locally maintains (i) a translator implementing the token-ID remapping, and (ii) the bijection specification (generated from a seed/configuration).

\paragraph{Honest execution of inference/fine-tuning.}
We assume the provider executes inference and fine-tuning faithfully on the received inputs (honest-but-curious at the boundary): the provider may log and analyze transmitted content, but does not intentionally tamper with outputs to break the translation mechanism. This assumption matches typical enterprise API settings where the primary concern is \emph{plaintext exposure in transit and logs}, rather than an active sabotage model.

\paragraph{Trusted client environment (necessary condition).}
\alienlm\ requires the client-side translator and its bijection specification to remain confidential. If the client runtime or local storage is compromised, any deterministic translation scheme is trivially exposed. We treat client compromise as out of scope and discuss it explicitly in Section~\ref{app:non-goals}.

\subsection{Observer Scenarios (O1--O3)}
\label{app:observer}

We evaluate robustness under three observer scenarios, consistent with Section~\ref{subsec:assumptions} in the main text. These scenarios reflect increasing levels of access beyond passive API-boundary observation.

\begin{itemize}
  \item \textbf{O1: Passive observation.}
    The observer sees only alien text exchanged with the API (prompts and responses), and knows the public tokenizer/vocabulary. The observer may apply corpus statistics (e.g., frequency matching), off-the-shelf LLM/MT systems for inverse translation, and general heuristic decipherment. The observer cannot query the client translator and does not access any aligned plaintext--alien pairs beyond what is implicitly observable.

  \item \textbf{O2: Limited leakage.}
    In addition to O1, the observer obtains a \emph{bounded} number of aligned plaintext--alien pairs (e.g., from accidental logs, shared snippets, or partial disclosure). The observer can train or prompt inverse models using these examples and attempt extrapolation to unseen mappings. This scenario captures realistic ``some leakage happens'' cases without granting unrestricted access.

  \item \textbf{O3: Model access.}
    In addition to O1, the observer gains access to the \emph{adapted model parameters} (e.g., via extraction, insider access, or artifacts), but still does not obtain the translator or bijection specification. The observer may attempt to infer the mapping by analyzing embedding/LM-head structure, nearest-neighbor correspondences, or other parameter-space signals.
\end{itemize}

These scenarios are intentionally practical: they focus on what an API provider or a third-party observer could plausibly obtain, rather than assuming oracle access to the client translator.

\subsection{Protection Scope and Explicit Non-goals}
\label{app:non-goals}

\paragraph{What \alienlm\ protects (scope).}
\alienlm\ is a \emph{text-level translation layer} that reduces exposure of \emph{human-readable plaintext} at the API boundary. Under O1--O3, the observer sees alienized strings that are not directly interpretable without the client-held mapping configuration. In this sense, \alienlm\ aims to mitigate ``plaintext in transit/logs'' risk independent of provider-side retention policies.

\paragraph{What \alienlm\ does not claim.}
\alienlm\ provides \emph{empirical robustness} against recovery attempts within the evaluated scenarios; it does not claim formal guarantees against all possible adversaries or side channels. The intent is deployable, controllable reduction of plaintext exposure in black-box API usage.

\paragraph{Explicit non-goals.}
The following are out of scope by design:

\begin{itemize}
  \item \textbf{Client compromise.}
    If the translator binary, bijection specification, or client memory is exposed, the mapping can be reconstructed and the alien text becomes readable. This assumption that the client environment remains secure is standard across key-based privacy schemes and falls outside our API-boundary threat model.

  \item \textbf{Translation-oracle access (active querying).}
    If an attacker can query the translator on arbitrary chosen inputs (i.e., obtains oracle access to the encode/decode interface), they can enumerate mappings and build a decipherment dictionary. Since this requires access to the client-side translator, it constitutes a form of client compromise and falls outside our threat model, which focuses on passive observation at the API boundary.

  \item \textbf{Unbounded aligned leakage.}
    Our evaluation considers bounded plaintext--alien leakage (O2). If an attacker obtains arbitrarily large aligned corpora with broad vocabulary coverage, recovery can become progressively easier. The practical threshold depends on coverage and query diversity; we therefore report results under bounded leakage budgets.

  \item \textbf{Metadata and side channels.}
    Message length, token count, timing, formatting conventions, and syntactic structure (e.g., punctuation positions) remain observable. These can leak coarse information (e.g., ``this looks like code''), and mitigating them would require orthogonal techniques such as padding or structure obfuscation.

  \item \textbf{Behavioral fingerprinting.}
    Because \alienlm\ preserves model functionality, response styles and refusal patterns may still reveal coarse task categories. This is orthogonal to token-level translation and is not addressed here.
\end{itemize}

%% file: sections/07_appendix_formalization.tex
\section{Formalization of Alien Language and Translator}
\label{app:formal}

This appendix provides a consolidated formal definition of the Alien Language and the client-side translator used by \alienlm. We emphasize two properties: (i) \emph{API compatibility}, which requires only the public tokenizer and vocabulary, and (ii) \emph{lossless round-trip translation} for authorized clients.

\subsection{Notation and Special-Token Handling}
\label{app:formal-notation}

Let the target model's public vocabulary be
\[
  \mathcal{V}=\{(v_k,i_k)\}_{k=1}^{|\mathcal{V}|},
\]
where $v_k$ is the token string and $i_k$ is its token ID. Let $\mathcal{S}\subset \mathcal{V}$ denote the set of special tokens that must remain unchanged (e.g., BOS/EOS/PAD and reserved tokens). Define the set of permutable (non-special) token IDs:
\[
  I=\{\, i_k \mid (v_k,i_k)\in \mathcal{V},\ (v_k,i_k)\notin \mathcal{S}\,\}.
\]
We denote the public tokenizer as $\tau_{\text{tgt}}$ with inverse $\tau_{\text{tgt}}^{-1}$, mapping between text strings and token-ID sequences.

\subsection{Alien Language as a Vocabulary-Level Bijection}
\label{app:formal-bijection}

\textbf{Toy example.} Suppose we want to pair a token \texttt{come} with a different-looking but semantically related token. For each candidate, we combine semantic closeness (embedding similarity) and surface dissimilarity (edit distance) into a score. The best trade-off wins (Table~\ref{tab:toy-bijection}). If the bijection includes \texttt{come}$\leftrightarrow$\texttt{world} and \texttt{here}$\leftrightarrow$\texttt{cup}, then plaintext \texttt{come here} becomes alien text \texttt{world cup}.
\begin{table}[t]
  \centering
  \caption{Toy example for bijection initialization (higher score is better).}
  \label{tab:toy-bijection}
  \begingroup
  \setlength{\tabcolsep}{5pt}
  \begin{tabular}{@{}lcccc@{}}
    \toprule
    Candidate & Embed sim & Sim dist & Edit dist & Score \\
    \midrule
    comes & 0.92 & 0.08 & 1 & 0.84 \\
    hello & 0.06 & 0.94 & 4 & 2.12 \\
    world & 0.40 & 0.60 & 4 & 2.80 \\
    cup   & 0.07 & 0.93 & 3 & 1.14 \\
    here  & 0.80 & 0.20 & 3 & 2.60 \\
    \bottomrule
  \end{tabular}
  \endgroup
\end{table}

\paragraph{Bijection definition.}
We define the Alien Language by a bijection over the non-special token ID set:
\[
  f: I \rightarrow I,
\]
which induces a deterministic remapping of token-ID sequences. Special tokens are fixed by construction (i.e., not permuted).

\subsection{Client-side Translator and Correctness}
\label{app:translator-correctness}

\paragraph{Translator definition.}
The client-side translator consists of an encoder $E$ (plaintext $\rightarrow$ alien text) and decoder $D$ (alien text $\rightarrow$ plaintext):
\begin{align}
  E(x) &= \tau_{\text{tgt}}^{-1}\!\left(f\!\left(\tau_{\text{tgt}}(x)\right)\right), \label{eq:app-enc}\\
  D(x') &= \tau_{\text{tgt}}^{-1}\!\left(f^{-1}\!\left(\tau_{\text{tgt}}(x')\right)\right). \label{eq:app-dec}
\end{align}
Special tokens are excluded from permutation, so they are preserved under both $E$ and $D$.

\paragraph{Lossless round-trip translation.}
For any text $x$ that is representable under $\tau_{\text{tgt}}$, the translator is lossless:
\[
  D(E(x)) = x.
\]
\emph{Proof sketch.} Let $z=\tau_{\text{tgt}}(x)$ be the token-ID sequence. Applying Eq.~\ref{eq:app-enc} yields $E(x)=\tau_{\text{tgt}}^{-1}(f(z))$. Re-tokenizing gives $\tau_{\text{tgt}}(E(x))=f(z)$ because the transformation is defined purely in the ID space under the same tokenizer and vocabulary. Applying Eq.~\ref{eq:app-dec} yields
$\tau_{\text{tgt}}^{-1}(f^{-1}(f(z)))=\tau_{\text{tgt}}^{-1}(z)=x$.

\paragraph{API-boundary property.}
Authorized users transmit only alien text $x'$ to the API and decode the received alien output $\hat{y}'$ locally:
\[
  x \xrightarrow{E} x' \xrightarrow{\text{API}} \hat{y}' \xrightarrow{D} \hat{y}.
\]
The API observes $x',\hat{y}'$ only, while the client recovers $x,\hat{y}$ via the translator.

\subsection{Alienization Ratio \texorpdfstring{$\rho$}{rho} (Partial Permutation)}
\label{app:rho}

\paragraph{Motivation.}
To control the opacity--utility trade-off, \alienlm\ supports \emph{partial} alienization, leaving a fraction of token IDs unchanged.

\paragraph{Definition.}
Let $\rho\in[0,1]$ and let $I_{\rho}\subseteq I$ be a subset with $|I_{\rho}|=\lfloor \rho|I|\rfloor$. We define a partial bijection $f_{\rho}$ as:
\[
  f_{\rho}(i)=
  \begin{cases}
    f(i), & i\in I_{\rho},\\
    i, & i\notin I_{\rho}.
  \end{cases}
\]
The corresponding translator is:
\begin{align}
  E_{\rho}(x) &= \tau_{\text{tgt}}^{-1}\!\left(f_{\rho}\!\left(\tau_{\text{tgt}}(x)\right)\right), \\
  D_{\rho}(x') &= \tau_{\text{tgt}}^{-1}\!\left(f_{\rho}^{-1}\!\left(\tau_{\text{tgt}}(x')\right)\right),
\end{align}
which still satisfies $D_{\rho}(E_{\rho}(x))=x$.

To illustrate, the encoder $E_{\rho}$ proceeds in three steps: (1) tokenize plaintext $x$ into a sequence of token IDs via $\tau_{\text{tgt}}$, (2) apply the partial bijection $f_{\rho}$ to remap IDs in $I_{\rho}$ while leaving others unchanged, and (3) convert the remapped IDs back to text via $\tau_{\text{tgt}}^{-1}$, producing alien text. The decoder $D_{\rho}$ reverses this process by applying $f_{\rho}^{-1}$ to recover the original IDs, ensuring lossless round-trip translation. Algorithm~\ref{alg:translator-min} provides the pseudocode.

\begin{algorithm}[h]
  \caption{Translator (encode/decode) with token-ID remapping.}
  \label{alg:translator-min}
  \begin{algorithmic}[1]
    \FUNCTION{\textsc{Encode}$(x)$}
      \STATE $z \leftarrow \tau_{\text{tgt}}(x)$
      \STATE $z' \leftarrow f_{\rho}(z)$ \COMMENT{apply ID remapping elementwise (skip special tokens)}
      \STATE \textbf{return} $\tau_{\text{tgt}}^{-1}(z')$
    \ENDFUNCTION
    \FUNCTION{\textsc{Decode}$(x')$}
      \STATE $z' \leftarrow \tau_{\text{tgt}}(x')$
      \STATE $z \leftarrow f_{\rho}^{-1}(z')$
      \STATE \textbf{return} $\tau_{\text{tgt}}^{-1}(z)$
    \ENDFUNCTION
  \end{algorithmic}
\end{algorithm}

\paragraph{Implementation note.}
In our experiments, $I_{\rho}$ is selected by a fixed randomized procedure derived from a client-held configuration (e.g., a seed), and is held constant within an experiment/run. This ensures consistent translation across prompts, responses, and training pairs.

%% file: sections/08_appendix_bijection.tex
\section{Bijection Optimization: Objective, Proxy Signals, and Approximate Solver}
\label{app:bijection}

This appendix expands Section~\ref{subsec:optimization} by detailing (i) the optimization objective used to construct the bijection, (ii) how we obtain proxy signals under black-box constraints, and (iii) the scalable approximate solver used at vocabulary scale.

\subsection{Objective and Design Trade-off}
\label{app:bijection-objective}

\paragraph{Setup.}
Let $I$ be the set of non-special token IDs (Appendix~\ref{app:formal-notation}). For each token ID $i\in I$, let $s(i)$ denote its surface string in the public vocabulary. We use the length-normalized edit distance
\[
  \tilde d_{\text{edit}}(a,b)=\frac{d_{\text{edit}}(a,b)}{\max(|a|,|b|)}.
\]
Let $e_{\star}(i)\in\mathbb{R}^{d}$ be a token representation used to estimate semantic relatedness (defined in Section~\ref{app:proxy}).

\paragraph{Objective.}
We construct a bijection over the active set $I_{\rho}\subseteq I$ that achieves high surface dissimilarity for human opacity while preserving semantic similarity for model learnability. The natural formulation imposes a hard constraint $d_{\text{sim}}(e_{\star}(i), e_{\star}(f(i))) \le \alpha$ for all $i$; relaxing this via a Lagrange multiplier yields the unconstrained objective:
\begin{equation}
  \label{eq:app-bijection-obj}
  \max_{f\in\mathfrak{S}(I_{\rho})}\quad
  \sum_{i\in I_{\rho}}
  \tilde d_{\text{edit}}\big(s(i), s(f(i))\big)\;-\;\mu \, d_{\text{sim}}\big(e_{\star}(i), e_{\star}(f(i))\big),
\end{equation}
where $\mathfrak{S}(I_{\rho})$ denotes the set of bijections on $I_{\rho}$, $d_{\text{sim}}$ is cosine distance, and $\mu \ge 0$ controls the trade-off between opacity and learnability. In black-box settings where target embeddings are unavailable, we instantiate $e_{\star}$ with proxy embeddings $e_P$ from an open-source model (Section~\ref{app:proxy}).

\paragraph{Interpretation.}
The first term pushes mapped token strings to look different from their originals, reducing readability of transmitted text. The second term discourages mapping semantically unrelated tokens, which empirically improves adaptation efficiency and downstream utility after AAT.

\subsection{Proxy Representations under Black-box Constraints}
\label{app:proxy}

\paragraph{Motivation.}
In black-box API settings, we cannot access target-model embeddings or LM-head parameters. We therefore approximate semantic similarity using proxy embeddings $e_{P}$ from an open-source model.

\paragraph{Token representations.}
If the proxy model shares the same vocabulary as the target, we directly use its token embedding or LM-head vectors. If vocabularies differ, we embed a target token string via proxy subpieces. Let $\tau_{\text{proxy}}$ be the proxy tokenizer, and let
\[
  S(i)=\tau_{\text{proxy}}(s(i))
\]
be the proxy subpiece sequence for the target token surface string. We define:
\[
  e_{P}(i)=\frac{1}{|S(i)|}\sum_{u\in S(i)} e_{P}(u),
\]
where $e_{P}(u)$ is the proxy vector for proxy token $u$ (we use LM-head vectors in the paper unless stated otherwise).

\paragraph{Practical note.}
This construction is used only to rank semantically plausible candidates during bijection search. The API-side model remains unchanged at construction time; learnability is validated by AAT results (Table~\ref{tab:main}) and ablation (Table~\ref{tab:ablation}).

\subsection{Approximate Solver: kNN Candidate Reduction with Greedy Pairing}
\label{app:solver}

\paragraph{Motivation}
Directly optimizing Eq.~\ref{eq:app-bijection-obj} over $|I_{\rho}|\approx 10^5$ tokens is computationally infeasible if we consider all $O(n^2)$ candidate pairs, and exact global assignment methods are intractable at this scale. We therefore use a scalable approximation.

\paragraph{Pair score.}
We compute a score for candidate pairs:
\[
  S(i,j)=\tilde d_{\text{edit}}\big(s(i),s(j)\big)\;-\;\mu\, d_{\text{sim}}\big(e_{P}(i),e_{P}(j)\big).
\]

\paragraph{kNN candidate reduction.}
The key observation is that most token pairs already exhibit high surface dissimilarity (the first term in $S(i,j)$ is naturally large), whereas semantically similar pairs are rare across the vocabulary. Filtering by edit distance would retain too many candidates, while filtering by embedding similarity effectively narrows the search to tokens where the learnability term matters. We therefore retrieve, for each token $i$, a candidate set $\mathcal{C}(i)$ consisting of the top-$k$ nearest neighbors of $e_P(i)$ under cosine similarity (via approximate nearest neighbor search), then evaluate the full score $S(i,j)$ only for $j\in\mathcal{C}(i)$.

\paragraph{Greedy symmetric pairing.}
We traverse tokens and greedily form disjoint pairs $(i,j)$:
(i) select the best available $j\in\mathcal{C}(i)$ maximizing $S(i,j)$,
(ii) set $f(i)=j$ and $f(j)=i$,
(iii) remove both from the available pool.
Any remaining tokens are paired arbitrarily as a fallback to complete a bijection. Algorithm~\ref{alg:app-bijection} provides the complete procedure.

\begin{algorithm}[t]
  \caption{Approximate bijection via kNN candidate reduction and greedy pairing.}
  \label{alg:app-bijection}
  \begin{algorithmic}[1]
    \INPUT $I_{\rho}$: IDs to permute;\; $s(\cdot)$: surface strings;\; $e_P(\cdot)$: proxy vectors;\;
    $k$: \#neighbors;\; $\mu$: trade-off weight
    \OUTPUT Bijection $f:I_{\rho}\to I_{\rho}$
    \STATE Build ANN index over $\{e_P(i)\}_{i\in I_{\rho}}$
    \STATE $Available\leftarrow I_{\rho}$
    \FORALL{$i\in I_{\rho}$}
      \IF{$i\notin Available$}
        \STATE \textbf{continue}
      \ENDIF
      \STATE $\mathcal{C}(i)\leftarrow \textsc{TopKNN}(e_P(i),k) \cap Available$
      \STATE $j^\star\leftarrow \arg\max_{j\in \mathcal{C}(i),\ j\neq i}\;\Big[
        \tilde d_{\text{edit}}(s(i),s(j))-\mu d_{\text{sim}}(e_P(i),e_P(j))
      \Big]$
      \IF{$j^\star$ exists}
        \STATE $f(i)\leftarrow j^\star;\ \ f(j^\star)\leftarrow i$
        \STATE $Available\leftarrow Available\setminus\{i,j^\star\}$
      \ENDIF
    \ENDFOR
    \STATE Pair remaining IDs in $Available$ arbitrarily to complete $f$
    \STATE \textbf{return} $f$
  \end{algorithmic}
\end{algorithm}

\subsection{Complexity}
\label{app:complexity}

Let $n=|I_{\rho}|$, embedding dimension $d$, average token-string length $\ell$, and neighbor count $k$.

\paragraph{Nearest-neighbor retrieval.}
We use FAISS~\citep{johnson2019billion} with an inner-product index on L2-normalized embeddings. Building the index requires $O(nd)$ time and space; querying $k$ neighbors for all $n$ tokens costs approximately $O(nk\log n)$.

\paragraph{Scoring.}
For each of $nk$ candidate pairs, edit distance costs $O(\ell^2)$ and cosine distance costs $O(d)$, yielding
\[
  O\big(nk(\ell^2+d)\big).
\]

\paragraph{Total.}
Overall, the solver runs in
\[
  O\big(nk(\ell^2+d+\log n)\big)
\]
time and uses $O(n + nk)$ memory for storing the ANN index and candidate sets. In practice, the full vocabulary build completes within minutes on a single machine (Appendix~\ref{app:env}).
\subsection{Bijection Hyperparameters}
\label{app:hp-bijection}
We use cosine distance for $d_{\text{sim}}$ with L2-normalized vectors and length-normalized Levenshtein distance for $\tilde d_{\text{edit}}$. Table~\ref{tab:hp-bijection} lists default hyperparameters.

\begin{table}[ht]
  \centering
  \caption{Hyperparameters for vocabulary bijection optimization.}
  \label{tab:hp-bijection}
  \begin{tabular}{@{}p{0.40\columnwidth}p{0.52\columnwidth}@{}}
    \toprule
    Setting & Value \\
    \midrule
    Neighbor count ($k$) & 100 \\
    Greedy batch size ($B$) & 50 \\
    Trade-off weight ($\mu$) & 1 (default) \\
    \bottomrule
  \end{tabular}
\end{table}

\subsection{AAT Hyperparameters}
\label{app:hp-train}
Table~\ref{tab:hp-train} lists the training hyperparameters used across all backbone models. The effective global batch size is computed as local\_bsz $\times$ grad\_acc $\times$ \#GPUs $= 2 \times 4 \times 4 = 32$. We enable sample packing to reduce padding overhead at a fixed maximum length of 2048 tokens. Mixed precision training with \texttt{bf16} improves memory efficiency without numerical instability. Also, we used Paged Adaw (8-bit)~\citep{dettmers2022bit} for efficient memory usage.

\begin{table}[ht]
  \centering
  \caption{AAT training hyperparameters (defaults across backbones).}
  \label{tab:hp-train}
  \begin{tabular}{@{}p{0.45\columnwidth}p{0.45\columnwidth}@{}}
    \toprule
    Setting & Value \\
    \midrule
    Global batch size & 32 \\
    Accumulation steps & 4 \\
    Local batch size & 2 \\
    Max seq. length & 2048 \\
    Optimizer & AdamW (8-bit) \\
    LR schedule & Constant \\
    Learning rate & 2e-5 \\
    Sample packing & True \\
    Mixed precision & Bfloat16  \\
    \bottomrule
  \end{tabular}
\end{table}

\subsection{Compute Environment and Training Cost}
\label{app:aat-cost}
\label{app:env}
Table~\ref{tab:env-aat} summarizes our compute environment. For open-source models (LLaMA~3~8B, Qwen~2.5-7B), we performed full-parameter fine-tuning on 4$\times$ NVIDIA A100-SXM4-80GB GPUs connected via NVLink. Each training run processed approximately 10M tokens over 3 epochs, completing in under 12 hours per model.

\begin{table}[ht]
  \centering
  \caption{Compute environment.}
  \label{tab:env-aat}
  \begin{tabular}{@{}p{0.34\columnwidth}p{0.58\columnwidth}@{}}
    \toprule
    Component & Spec / Notes \\
    \midrule
    GPU & NVIDIA A100-SXM4-80GB $\times$ 4 (NVLink) \\
    CPU & AMD EPYC 7763 64-Core $\times$ 2 \\
    Memory & 2.0 TiB \\
    AAT training time & $<$12 hours per model \\
    Bijection build time & $\leq$20 minutes \\
    \bottomrule
  \end{tabular}
\end{table}

Using LLaMA~3~8B as a reference, each AAT run consists of $\sim$8.3k optimization steps. Under on-demand cloud pricing of
\$2-4 per A100 GPU-hour (e.g., AWS p4d), the total cost scales with the step throughput
(steps/sec) and is roughly on the order of \$100--150 for our setup; with owned on-prem
hardware, the marginal cost is dominated by electricity.

% \vspace{-2.5em}
\cradd{\subsection{Serving-Time Overhead}}
\cradd{\label{app:serving-overhead}}
% \vspace{-2.3em}

\cradd{To assess deployment feasibility, we measured both the client-side translator overhead and end-to-end serving performance. Table~\ref{tab:serving-e2e} compares original and AlienLM pipelines on LLaMA~3~8B; Table~\ref{tab:serving-translator} breaks down the translator cost.}

\cradd{\begin{table}[h]
  \centering
  \caption{End-to-end serving comparison (LLaMA~3~8B).}
  \label{tab:serving-e2e}
  \begin{tabular}{@{}l c c@{}}
    \toprule
    Metric & Original & AlienLM \\
    \midrule
    Latency (s)       & 62.13  & 61.82 \\
    Throughput (tok/s) & 34.15  & 34.39 \\
    GPU memory (GB)    & 15.32  & 15.32 \\
    \bottomrule
  \end{tabular}
\end{table}}

\cradd{\begin{table}[ht]
  \centering
  \caption{Client-side translator overhead.}
  \label{tab:serving-translator}
  \begin{tabular}{@{}l l r@{}}
    \toprule
    Stage & Unit & Time (ms) \\
    \midrule
    Alienization & per sample & 0.21 \\
    Alienization & per token  & $<$0.01 \\
    Recovery     & per sample & 0.35 \\
    Recovery     & per token  & 0.01 \\
    \midrule
    \multicolumn{2}{@{}l}{Translator memory (MB)} & 22.23 \\
    \bottomrule
  \end{tabular}
\end{table}}

\cradd{The translator operates on CPU via dictionary lookup and adds sub-millisecond latency per sample. End-to-end latency and throughput remain within measurement noise, confirming that the dominant cost is remote model inference and network transfer, not local translation.}

\cradd{\subsection{Token Count and API Cost}}
\cradd{\label{app:token-cost}}

\cradd{AlienLM preserves the number of token IDs per sample by construction (the bijection is a one-to-one mapping within the same vocabulary). However, because the alien tokenizer maps to different surface strings, a third-party tokenizer used for billing may count tokens differently. Table~\ref{tab:token-inflation} reports tokenizer-level token count differences on a 10K-sample subset of the training pool.}

\cradd{The total training corpus comprises approximately 279M tokens under \texttt{o200k\_base} (450K samples). Because commercial fine-tuning APIs may apply their own chat formatting or internal tokenization, billed token counts can differ from the raw counts above; we report these figures as practical reference points rather than exact billing estimates.}

\cradd{\begin{table}[h]
  \centering
  \caption{Token count comparison across tokenizers on the same 10K-sample subset (plain text, no chat template).}
  \label{tab:token-inflation}
  \begin{tabular}{@{}l r r@{}}
    \toprule
    Tokenizer & Avg tokens/sample & Relative to LLaMA \\
    \midrule
    LLaMA~3 & 617.81 & --- \\
    Qwen~2.5 & 626.73 & +1.44\% \\
    Gemma~2  & 644.68 & +4.35\% \\
    \bottomrule
  \end{tabular}
\end{table}}

%% file: sections/09_appendix_additional_experiments.tex
\section{Additional Experiments and Ablations}
\label{app:extra-exp}

This appendix provides extended results and ablations that complement Section~\ref{subsec:adaptation}--\ref{subsec:seed}. We focus on additional quantitative tables and controlled variations, while keeping the main text concise.

% \begin{table*}[t]
%   \centering
%   \caption{Ablations on LLaMA~3~8B (accuracy, \%). \textsc{Average} is the unweighted mean over seven benchmarks. ${}^{\dagger}$ uses proxy head $e_P$ under the black-box constraint. \cradd{${}^{\ddagger}$ uses full white-box access (gradient-based embedding alignment); included as an upper-bound reference, not a deployable baseline.}}
%   \label{tab:ablation}
%   \adjustbox{width=\textwidth}{
%     \begin{tabular}{llcccccccc}
%       \toprule
%       \textbf{Methods}& \textbf{Components}& MMLU & ARC-E & ARC-C & HellaS & WinoG & TQA & GSM8K & Average \\
%       \midrule
%       \multirow{3}{*}{\makecell{LLaMA~3 \\ 8B}}
%       & Oracle & 67.32 & 84.13 & 59.39 & 57.07 & 74.35 & 35.25 & 75.89 & 64.77 \\
%       & SFT & 63.74 & 80.56 & 53.67 & 53.70 & 71.74 & 37.58 & 76.12 & 62.44 \\
%       & \cradd{SentinelLM (white-box)${}^{\ddagger}$} & \cradd{63.67} & \cradd{80.77} & \cradd{53.50} & \cradd{54.16} & \cradd{72.61} & \cradd{37.58} & \cradd{74.45} & \cradd{62.39} \\
%       \midrule
%       \multirow{4}{*}{AlienLM}
%       & $e_P$ LM Head${}^{\dagger}$ & 49.42 & 72.14 & \textbf{44.28} & 47.86 & 61.48 & 35.01 & 63.08 & 53.32 \\
%       & $e_{\text{tgt}}$ LM Head & \textbf{51.60} & \textbf{73.73} & 44.20 & \textbf{48.38} & \textbf{65.11} & \textbf{36.96} & \textbf{65.50} & \textbf{55.07} \\
%       & $e_{\text{tgt}}$ Embeddings & 50.82 & 68.64 & 43.67 & 47.98 & 64.01 & 36.47 & 64.14 & 53.68 \\
%       & Random $\mathcal{V}$ & 29.92 & 46.34 & 27.56 & 38.47 & 55.09 & 30.23 & 31.08 & 36.96 \\
%       \bottomrule
%     \end{tabular}
%   }
% \end{table*}

\subsection{Proxy vs.\ Target Representations (Ablation)}
\label{app:ablation}

Table~\ref{tab:ablation} isolates the effect of (i) proxy vs.\ target representations used for bijection construction and (ii) random permutation. This directly supports the black-box feasibility claim in the main text (Section~\ref{subsec:optimization}).

\paragraph{Key takeaways.}
Using proxy LM-head vectors ($e_P$) performs close to using target LM-head vectors ($e_{\text{tgt}}$), with only a modest average gap, while random permutation severely degrades performance. This validates that proxy-based similarity structure is sufficient for practical bijection optimization under API-only constraints. \cradd{SentinelLM (white-box) serves as an upper-bound ceiling: with full gradient-based embedding alignment it achieves 62.39 average, close to the SFT reference (62.44). AlienLM's API-only proxy approach recovers 53.32, closing roughly 64\% of the gap between random bijection (36.96) and the white-box ceiling.}

\subsection{Alienization Ratio: Full Results}
\label{app:ratio}

\begin{table*}[t]
  \centering
  \caption{Effect of the alienization ratio $\rho$ on benchmarks (accuracy, \%).
  \textsc{Average} is the unweighted mean; Ratio is relative to Oracle.}
  \label{tab:alienization-ratio}
  \adjustbox{width=\textwidth}{
    \begin{tabular}{lccccccccc}
      \toprule
      \textbf{Method} & \textbf{Ratio (\%)} & \textbf{MMLU} & \textbf{ARC-E} & \textbf{ARC-C} & \textbf{HellaS} & \textbf{WinoG} & \textbf{TQA} & \textbf{GSM8K} & \textbf{Average} \\
      \midrule
      Oracle & 100   & 67.32 & 84.13 & 59.39 & 57.07 & 74.35 & 35.25 & 75.89 & 64.77 \\
      \midrule
      AlienLM ($\rho{=}0.2$)  & 93.06 & 60.18 & 77.61 & 52.05 & 53.32 & 70.01 & 37.58 & 71.19 & 60.28 \\
      AlienLM ($\rho{=}0.4$)  & 89.01 & 57.31 & 76.01 & 47.44 & 51.63 & 66.38 & 34.76 & 70.05 & 57.65 \\
      AlienLM ($\rho{=}0.6$)  & 85.93 & 53.98 & 74.33 & 44.62 & 49.70 & 65.43 & 35.74 & 65.81 & 55.66 \\
      AlienLM ($\rho{=}0.8$)  & 84.46 & 51.98 & 73.70 & 44.54 & 48.96 & 63.14 & 34.52 & 66.11 & 54.71 \\
      AlienLM ($\rho{=}1$) & 82.33 & 49.42 & 72.14 & 44.28 & 47.86 & 61.48 & 35.01 & 63.08 & 53.32 \\
      \bottomrule
    \end{tabular}
  }
\end{table*}

Table~\ref{tab:alienization-ratio} reports full benchmark scores across alienization ratios $\rho$, expanding Figure~\ref{fig:encratio_curve} in the main text (Section~\ref{subsec:alienizationratio}).

\paragraph{Consistent improvement across tasks.}
Performance improves monotonically as $\rho$ decreases across all seven benchmarks, with no exceptions. This consistency suggests that the opacity-utility trade-off is smooth and predictable, reducing the fraction of alienized tokens uniformly benefits task performance regardless of task type.

\paragraph{Task-specific observations.}
The largest absolute gains from $\rho=1.0$ to $\rho=0.2$ appear in knowledge-intensive tasks: MMLU (+10.76), WinoGrande (+8.53), and GSM8K (+8.11). These tasks likely benefit most from preserving original tokens that carry domain-specific or numerical semantics. In contrast, TruthfulQA shows relatively modest variation (+2.57), suggesting that truthfulness evaluation is less sensitive to surface-level token changes.

\paragraph{Practical implications.}
At $\rho=0.6$, AlienLM still alienizes the majority of tokens while achieving 86\% of Oracle performance. This demonstrates that selective alienization can provide substantial opacity with limited utility cost, enabling deployment strategies that balance protection requirements against performance constraints.

\paragraph{Note on interpretation.}
As $\rho$ decreases, fewer tokens are permuted and the model can rely more on original lexical anchors, improving utility. This table is intended as a deployment reference for selecting $\rho$.

\subsection{Domain-specific Adaptation: Full Results}
\label{app:domain}

\begin{table}[h!]
  \centering
  \caption{Domain-specific fine-tuning on code/math benchmarks (LLaMA~3~8B).
  \textsc{Average} over MBPP and HumanEval.}
  \label{tab:finetune-code}
  \adjustbox{width=\columnwidth}{
    \begin{tabular}{lcc|cc|c}
      \toprule
      \textbf{Method} & \textbf{Tokenizer} & \textbf{Data} &
      \textbf{MBPP} & \textbf{HumanEval} & \textbf{Average (Code)} \\
      \midrule
      full           & O    & O    & 27.25 & 47.56& 37.41\\
      - tokenizer    & X    & O    & 26.46 & 45.73& 36.10\\
      - data         & O    & X    & 20.63 & 29.27& 24.95\\
      - tok \& data  & X    & X    & 19.58 & 30.49& 25.04\\
      code/math only & only & only & 32.54 & 40.24& 36.39\\
      + data         & O    & +150k& 35.19 & 46.34& 40.77\\
      \bottomrule
    \end{tabular}
  }
\end{table}

Tables~\ref{tab:finetune-general} and \ref{tab:finetune-code} provide full results for the domain adaptation study in Section~\ref{subsec:domain}, including both general benchmarks and code/math benchmarks.

\subsection{Seed Robustness and Diversity}
\label{app:seed-diversity}

Table~\ref{tab:seed-diversity} reports robustness across multiple random seeds under the bucketed greedy solver (Section~\ref{subsec:seed}). This supports per-user key diversification without large variance in utility.

\begin{figure}
  \centering
  \includegraphics[width=\columnwidth]{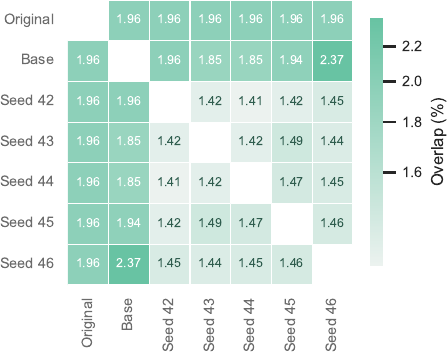}
  \caption{Pairwise token-overlap heatmap between seeds (overlap ratio, \%; lower is better).}
  \label{fig:tokrob2}
\end{figure}

\paragraph{Summary.}
Across seeds, performance variance is small, while overlap between seeds remains low (Figure~\ref{fig:tokrob2}). This enables issuing distinct bijections for different users/sessions without materially changing model utility.

\begin{table}
  \centering
  \caption{Multi-tenant vs.\ per-key adaptation (LLaMA~3~8B).}
  \label{tab:tenant}
  \begin{tabular}{lcccc}
    \toprule
    Setting & Avg & Ratio & MMLU & GSM8K \\
    \midrule
    Per-key \\ (avg over 5 seeds) & 50.95 & 78.67 & 45.67 & 59.90 \\
    Tenant \\ (avg over 5 seeds) & 41.26 & 63.69 & 28.39 & 37.86 \\
    \bottomrule
  \end{tabular}
\end{table}

\subsection{Training Hyperparameters and Cost}
\label{app:train-details}

For reproducibility, Table~\ref{tab:hp-train} lists default AAT hyperparameters that correspond to Section~\ref{subsec:adaptation} in the main text.
Unless otherwise noted, all methods in Table~\ref{tab:main} are trained with the same AAT budget (epochs, steps, and batch configuration).

\subsection{Training Data Details}
\label{app:train-data}
We use Magpie instruction-tuning data \citep{magpie} with 300K filtered instruction pairs, and for domain-focused AAT we sample 150K math/coding-annotated examples.\footnote{\texttt{Magpie-Align/Magpie-Pro-300K-Filtered}; \texttt{Magpie-Align/Magpie-Llama-3.1-Pro-300K-Filtered}; \texttt{Magpie-Align/Magpie-Llama-3.3-Pro-500K-Filtered}.} These datasets cover coding, math, and general Q\&A and are used consistently across backbones.

\cradd{\subsection{Data-Volume Ablation}}
\cradd{\label{app:data-volume}}

\cradd{To understand how much adaptation data is needed, we trained LLaMA~3~8B AlienLM variants on 50K, 150K, and the full 450K training pool (evaluated at both 0.5 and 1 epoch checkpoints). All subset runs use gradient accumulation 2 and are step-matched to the full 1-epoch checkpoint for fair comparison. Table~\ref{tab:data-volume} reports results on the main evaluation suite.}

\cradd{\begin{table}
  \centering
  \caption{Data-volume ablation (LLaMA~3~8B AlienLM, $\rho{=}1.0$).}
  \label{tab:data-volume}
  \adjustbox{width=\columnwidth}{
    \begin{tabular}{l c c c c}
      \toprule
      Task & 50K & 150K & Full (0.5 ep.) & Full (1 ep.) \\
      \midrule
      MMLU          & 35.76 & 39.56 & 35.97 & 39.20 \\
      ARC-Easy      & 59.72 & 65.03 & 62.54 & 64.23 \\
      ARC-Challenge & 39.76 & 44.80 & 41.81 & 45.48 \\
      HellaSwag     & 52.78 & 59.60 & 59.51 & 60.89 \\
      WinoGrande    & 56.91 & 60.06 & 59.59 & 62.04 \\
      TruthfulQA    & 32.56 & 35.13 & 35.50 & 35.13 \\
      GSM8K (CoT)   & 40.56 & 50.34 & 45.11 & 55.65 \\
      \midrule
      Macro Avg     & 45.43 & 50.64 & 48.58 & 51.80 \\
      \bottomrule
    \end{tabular}
  }
\end{table}}

\cradd{The 150K subset recovers most of the full-data performance, trailing the full 1-epoch checkpoint by only 1.16 macro-average points. In contrast, 50K samples are insufficient for stable adaptation across all tasks. GSM8K remains the most budget-sensitive benchmark, consistent with the observation that mathematically structured tasks are most sensitive to adaptation budget.}

\cradd{\subsection{Generation Evaluation}}
\cradd{\label{app:generation}}

\cradd{To verify that the performance gap observed in multiple-choice benchmarks also extends to free-form generation, we evaluate TruthfulQA generation and two LongBench tasks.}

\cradd{\subsubsection{TruthfulQA Generation}}
\cradd{\label{app:truthfulqa-gen}}

\cradd{We evaluate free-form generation on TruthfulQA~\citep{truthfulqa} using BLEU Acc and ROUGE-L Acc, which measure whether the generated answer matches the reference truthful answer set. Table~\ref{tab:truthfulqa-gen} reports results across three backbones.}

\cradd{\begin{table}[h]
  \centering
  \caption{TruthfulQA generation results (higher is better).}
  \label{tab:truthfulqa-gen}
  \begin{tabular}{@{}l cc cc@{}}
    \toprule
    \multirow{2}{*}{Model} & \multicolumn{2}{c}{BLEU Acc} & \multicolumn{2}{c}{ROUGE-L Acc} \\
    \cmidrule(lr){2-3} \cmidrule(lr){4-5}
    & Original & AlienLM & Original & AlienLM \\
    \midrule
    LLaMA~3~8B  & 0.463 & 0.393 & 0.475 & 0.404 \\
    Qwen~2.5~7B & 0.506 & 0.382 & 0.502 & 0.393 \\
    Gemma~2~9B  & 0.490 & 0.416 & 0.515 & 0.441 \\
    \bottomrule
  \end{tabular}
\end{table}}

\cradd{All three backbones degrade under AlienLM, confirming that the performance gap is not confined to multiple-choice evaluation. The degradation pattern is backbone-dependent: Qwen shows the largest drop alongside a notable shortening of generated answers, while LLaMA and Gemma lose accuracy despite producing answers of similar length, suggesting a broader reduction in free-form answer quality rather than a simple length effect.}

\cradd{\subsubsection{LongBench}}
\cradd{\label{app:longbench}}

\cradd{We evaluate two long-context generation tasks from LongBench: \texttt{gov\_report\_e} (summarization) and \texttt{qasper\_e} (question answering over papers). Table~\ref{tab:longbench} reports F1 scores.}

\cradd{\begin{table}[h]
  \centering
  \caption{LongBench results (F1, higher is better).}
  \label{tab:longbench}
  \begin{tabular}{@{}l cc cc@{}}
    \toprule
    \multirow{2}{*}{Model} & \multicolumn{2}{c}{\texttt{gov\_report\_e}} & \multicolumn{2}{c}{\texttt{qasper\_e}} \\
    \cmidrule(lr){2-3} \cmidrule(lr){4-5}
    & Original & AlienLM & Original & AlienLM \\
    \midrule
    LLaMA~3~8B  & 0.281 & 0.228 & 0.128 & 0.248 \\
    Qwen~2.5~7B & 0.317 & 0.156 & 0.108 & 0.061 \\
    Gemma~2~9B  & 0.256 & 0.188 & 0.169 & 0.080 \\
    \bottomrule
  \end{tabular}
\end{table}}

\cradd{\texttt{gov\_report\_e} degrades consistently across all backbones. \texttt{qasper\_e} shows mixed behavior: LLaMA improves substantially while Qwen and Gemma degrade. This task-dependent pattern indicates that long-context generation under AlienLM is not uniformly degraded but varies with both the task and the backbone model.}

%% file: sections/10_appendix_recovery.tex
\section{Recovery Robustness: Settings and Extended Results}
\label{app:recovery}

This appendix provides extended details for Section~\ref{subsec:robustness}. We evaluate whether an observer can recover token mappings or reconstruct readable plaintext from alien text under three observer scenarios: O1 (passive observation), O2 (limited leakage of plaintext--alien pairs), and O3 (access to the adapted model weights \cradd{while the client-held translator and bijection remain unavailable}).

\subsection{Common Setup and Metrics}
\label{app:recovery-setup}

\paragraph{Data.}
Unless otherwise noted, attacks use held-out alienized prompts and responses from our evaluation suite. The observer always knows the public tokenizer and vocabulary of the target model.

\paragraph{What counts as ``recovery.''}
We report recovery at two granularities:
\begin{itemize}
  \item \textbf{Token recovery (mapping accuracy).} The fraction of token IDs correctly mapped back to their original IDs for the tokens appearing in the evaluation corpus. This measures whether the observer can infer the bijection entries.
  \item \textbf{Text recovery (readability).} For learning-based decoders (LLM/MT), we measure BLEU and ROUGE-L against the reference plaintext, and additionally report an LLM-based judge score (1--3) for overall readability. \cradd{The judge (GPT-5.1) rates each decoded output on a 1--3 integer scale (1\,=\,no correspondence to the reference, 3\,=\,perfect match) across three dimensions: semantic fidelity, structural consistency, and overall quality. We report the overall dimension in all tables.}
\end{itemize}
We emphasize that text recovery can remain low even if a small fraction of frequent tokens are inferred.

\subsection{O1: Passive Observation (Frequency Matching)}
\label{app:o1-frequency}

\paragraph{Attack.}
The observer collects token-frequency statistics from alien text and attempts to match them to public corpora token frequencies under the same tokenizer. This is the natural analogue of classical frequency analysis, but applied to subword token IDs rather than characters.

\paragraph{Result.}
As summarized in Table~\ref{tab:robustness}, frequency matching recovers fewer than 0.01\% of tokens. In practice, only a tiny handful of extremely frequent tokens have consistent frequency signatures; beyond that head, subword token frequency spectra are \cradd{far flatter than character-level distributions}, and the mapping is underdetermined.

\paragraph{Implementation note.}
We perform matching on the top-$m$ most frequent tokens and report recovery over the full evaluation vocabulary. We also tested alternative public corpora sources and observed similarly negligible recovery.

\subsection{O2: \cradd{Text and Token} Recovery with Limited Leakage}
\label{app:o2-learning}

\cradd{Under O2, the observer holds a bounded number of plaintext--alien pairs. We evaluate two attack goals: (i) \emph{text recovery}, where the observer attempts to decode unseen alien text into readable plaintext, and (ii) \emph{token recovery}, where the observer attempts to reconstruct bijection entries beyond the leaked pairs.}

\cradd{\subsubsection{Text recovery: parallel few-shot inverse translation}}
\label{app:o2-parallel}

\paragraph{Attack.}
The observer receives up to $n$ aligned plaintext--alien pairs (shots), and prompts an LLM to translate alien text back to plaintext. We test $n\in\{0,1,5,20\}$.

\paragraph{Result.}
Table~\ref{tab:llm-attack-parallel} shows that even with 20 parallel examples, inverse-translation remains poor (BLEU $<12$, low judge scores), indicating that limited aligned leakage is insufficient for decipherment at vocabulary scale.

\begin{table}[ht]
  \centering
  \caption{LLM inverse-translation with parallel plaintext--alien text leakage. LLM-Judge: 1-3 (higher is better).}
  \label{tab:llm-attack-parallel}
  \adjustbox{width=\columnwidth}{
    \begin{tabular}{l c c c c}
      \toprule
      Model & Shots & BLEU & ROUGE-L & LLM-Judge (Overall) \\
      \midrule
      GPT-5.1 & 0  & 3.22  & 0.16 & 1.07 \\
      GPT-5.1 & 1  & 6.16  & 0.19 & 1.16 \\
      GPT-5.1 & 5  & 9.50  & 0.23 & 1.16 \\
      GPT-5.1 & 20 & 11.56 & 0.25 & 1.54 \\
      \midrule
      GPT-5-mini & 0  & 1.46 & 0.14 & 1.04 \\
      GPT-5-mini & 1  & 0.79 & 0.11 & 1.04 \\
      GPT-5-mini & 5  & 2.17 & 0.15 & 1.11 \\
      GPT-5-mini & 20 & 1.88 & 0.15 & 1.17 \\
      \midrule
      GPT-4.1 & 0  & 1.19 & 0.11 & 1.01 \\
      GPT-4.1 & 1  & 2.63 & 0.14 & 1.06 \\
      GPT-4.1 & 5  & 5.82 & 0.19 & 1.11 \\
      GPT-4.1 & 20 & 4.64 & 0.20 & 1.11 \\
      \bottomrule
    \end{tabular}
  }
\end{table}

\cradd{\subsubsection{Text recovery: non-parallel inverse translation}}
\label{app:o2-nonparallel}

\paragraph{Attack.}
The observer is given a plaintext corpus and an alien corpus without alignment and attempts to infer a translation rule or mapping implicitly (e.g., by inducing a decoder with weak supervision).

\paragraph{Result.}
Table~\ref{tab:llm-attack-nonparallel} shows similarly low recoverability. Non-parallel signals provide little constraint because the transformation is a token-level relabeling rather than a natural language shift with shared substructure.

\begin{table}[ht]
  \centering
  \caption{LLM inverse-translation with non-parallel plaintext and alien corpora. LLM-Judge: 1-3 (higher is better).}
  \label{tab:llm-attack-nonparallel}
  \adjustbox{width=\columnwidth}{
    \begin{tabular}{l c c c c}
      \toprule
      Model & Shots & BLEU & ROUGE-L & LLM-Judge (Overall) \\
      \midrule
      GPT-5.1 & 1  & 6.31  & 0.19 & 1.22 \\
      GPT-5.1 & 5  & 9.74  & 0.24 & 1.45 \\
      GPT-5.1 & 20 & 10.18 & 0.23 & 1.61 \\
      \midrule
      GPT-5-mini & 1  & 0.86 & 0.11 & 1.01 \\
      GPT-5-mini & 5  & 2.08 & 0.15 & 1.08 \\
      GPT-5-mini & 20 & 1.86 & 0.16 & 1.13 \\
      \midrule
      GPT-4.1 & 1  & 1.82 & 0.13 & 1.03 \\
      GPT-4.1 & 5  & 4.00 & 0.18 & 1.15 \\
      GPT-4.1 & 20 & 4.13 & 0.19 & 1.08 \\
      \bottomrule
    \end{tabular}
  }
\end{table}

\FloatBarrier
\cradd{\subsubsection{Text recovery: MT-based decoding (NLLB)}}
\label{app:o2-mt}
\paragraph{Attack.}
We treat alien text as a source ``language'' and fine-tune a large MT foundation model, NLLB-200-3.3B~\citep{nllb2022} to translate it into English. The model is trained on SlimOrca~\citep{slimOrca} with alienized inputs as source and original English as target.

\paragraph{Training configuration.}
We fine-tune NLLB-200-3.3B for 2 epochs with batch size 16 (per-device batch size 4 $\times$ gradient accumulation 4), learning rate 5e-5, warmup steps 500, and maximum sequence length 1024 for both source and target. Training uses bfloat16 mixed precision. Training ran for 3,226 steps with a final loss of 2.87.

\paragraph{Result.}
As shown in Table~\ref{tab:mt-attack}, MT decoding fails (BLEU $<12$). NLLB-200-3.3B is a massively multilingual model trained on 200 languages, and recent work demonstrates that fine-tuned NLLB can effectively adapt to low-resource languages, even outperforming LLMs on such translation tasks~\citep{nllb-finetune}. Despite this strong cross-lingual transfer capability, the model fails to learn a coherent mapping from Alien Language to English. This suggests that Alien Language lies outside the distribution of human languages that multilingual MT models can exploit for transfer.

\begin{table}[ht]
  \centering
  \caption{MT-based inverse translation with NLLB-200-3.3B. LLM-Judge: 1-3 (higher is better).}
  \label{tab:mt-attack}
  \begin{tabular}{l c c c}
    \toprule
    BLEU & ROUGE-L & LLM-Judge (Overall) \\
    \midrule
    11.40 & 0.29 & 1.00 \\
    \bottomrule
  \end{tabular}
\end{table}

\FloatBarrier
\cradd{\subsubsection{Token recovery: known-plaintext leakage via n-gram extrapolation}}
\label{app:o2-ngram}

\paragraph{Attack.}
The observer receives up to 1{,}000 aligned plaintext--alien pairs and attempts to expand the mapping beyond seen tokens using n-gram co-occurrence statistics (e.g., hypothesizing that an unseen alien token corresponds to a plaintext token that frequently co-occurs with already matched neighbors).

\paragraph{Result.}
Table~\ref{tab:ngram-leakage} reports that token-level accuracy stays below 0.22\% and bijection-level accuracy remains 0\%. In other words, observed pairs do not extrapolate to unseen tokens: learning local phrase correspondences does not reveal the global vocabulary relabeling.

\begin{table}[ht]
  \centering
  \caption{Known-plaintext leakage via n-gram frequency analysis.}
  \label{tab:ngram-leakage}
  \adjustbox{width=\columnwidth}{
    \begin{tabular}{r r c c c}
      \toprule
      \# Pairs & \# Known Tokens & N-gram & Token Acc & Bijection Acc \\
      \midrule
      10   & 972   & 2 & 0.21\% & 0.00\% \\
      10   & 972   & 3 & 0.22\% & 0.00\% \\
      10   & 972   & 4 & 0.21\% & 0.00\% \\
      \midrule
      50   & 3223  & 2 & 0.19\% & 0.00\% \\
      50   & 3223  & 3 & 0.19\% & 0.00\% \\
      50   & 3223  & 4 & 0.18\% & 0.00\% \\
      \midrule
      1000 & 19673 & 2 & 0.19\% & 0.00\% \\
      1000 & 19673 & 3 & 0.19\% & 0.00\% \\
      1000 & 19673 & 4 & 0.18\% & 0.00\% \\
      \bottomrule
    \end{tabular}
  }
\end{table}

\FloatBarrier
\cradd{\subsubsection{Token recovery: collaborative known-plaintext attack}}
\cradd{\label{app:o2-collaborative}}

\cradd{\paragraph{Attack.}
The observer pools leaked plaintext--alien pairs from multiple users and applies the same n-gram plus Hungarian matching pipeline as in Section~\ref{app:o2-ngram}. We compare two settings: (i) \emph{shared seed}, where all users use the same bijection seed, and (ii) \emph{mixed seeds}, where users hold three distinct seeds (42/43/44). Only high-consensus token correspondences are retained (min\_confidence\,=\,0.95, min\_occurrences\,=\,3).}

\cradd{\paragraph{Result.}
Table~\ref{tab:collaborative-attack} shows that at moderate leakage budgets ($k$\,=\,1K and 5K), mixed-seed pooling yields consistently lower token accuracy than shared-seed pooling (e.g., 4.40\% vs.\ 6.05\% at $k$\,=\,1K). At a very large budget ($k$\,=\,10K), the gap largely closes. We conclude that per-user seeds materially reduce collusive leakage at low-to-moderate budgets but do not eliminate the risk under very large leakage.}

\FloatBarrier
\cradd{\subsubsection{Token recovery: observer tokenizer mismatch}}
\cradd{\label{app:o2-tokenizer}}

\cradd{\paragraph{Attack.}
We test whether the collaborative reconstruction attack depends on the observer knowing the victim tokenizer. Using the same LLaMA AlienLM plaintext--alien pairs ($k$\,=\,10K), we re-run the attack while tokenizing both sides with five common LLM tokenizers.}

\cradd{\paragraph{Result.}
Table~\ref{tab:tokenizer-mismatch} shows that the exact victim tokenizer (LLaMA~3) achieves 1.19\% accuracy, while all non-victim tokenizers fall to 0.05--0.11\%. This indicates that the attack is not tokenizer-invariant and becomes substantially less effective when the observer does not know the victim tokenization scheme.}

\cradd{\begin{table}
  \centering
  \caption{Collaborative known-plaintext attack: shared vs.\ mixed seeds.}
  \label{tab:collaborative-attack}
  \begin{tabular}{r c c}
    \toprule
    $k$ (pooled pairs) & Shared Seed & Mixed Seeds \\
    \midrule
    1{,}000  & 6.05\% & 4.40\% \\
    5{,}000  & 4.77\% & 3.19\% \\
    10{,}000 & 2.15\% & 2.56\% \\
    \bottomrule
  \end{tabular}
\end{table}}

\cradd{\begin{table}
  \centering
  \caption{Token recovery under observer tokenizer mismatch ($k$\,=\,10K).}
  \label{tab:tokenizer-mismatch}
  \begin{tabular}{l c}
    \toprule
    Observer Tokenizer & Token Acc \\
    \midrule
    LLaMA~3 (victim) & 1.19\% \\
    Qwen~2.5         & 0.05\% \\
    Mistral           & 0.07\% \\
    Gemma~2           & 0.08\% \\
    Phi-3             & 0.11\% \\
    \bottomrule
  \end{tabular}
\end{table}}

\FloatBarrier
\subsection{O3: Weight-based Mapping without the Bijection Seed}
\label{app:o3-weights}

\paragraph{Attack.}
The observer obtains the adapted model weights and attempts to recover mappings by comparing representations of alien tokens to original tokens. A simple instance is nearest-neighbor matching in embedding space (or LM-head space), reporting top-1 mapping accuracy. We report top-1 because bijection recovery requires exactly one correct mapping per token; even top-3 matching would yield $3^{|I_\rho|}$ candidate bijections, far too many to enumerate without additional constraints.

\paragraph{Result.}
As summarized in Table~\ref{tab:robustness}, this attack achieves $<0.11\%$ top-1 accuracy. Intuitively, the adapted model must encode many alien tokens in a way that supports next-token prediction, and representation neighborhoods can be ambiguous at scale. Without the seed, similarity alone does not uniquely identify the intended inverse mapping.

\paragraph{Details.}
We evaluate matching using cosine similarity on L2-normalized vectors. We report top-1 accuracy over the set of permuted non-special tokens. We also tested top-$k$ recovery (not shown) and found that while the true match can appear among multiple plausible candidates for some tokens, this does not translate into reliable end-to-end decipherment.

\subsection{Summary}
\label{app:recovery-summary}

Across O1--O3, recovery remains negligible under our evaluation: frequency-based heuristics fail beyond a tiny head; \cradd{under O2, text recovery (learning-based decoders), single-user token recovery (n-gram extrapolation), and collaborative token recovery all fail to generalize from bounded leakage, with per-user seeds further reducing collusive risk at moderate budgets and tokenizer mismatch degrading the attack substantially;} and weight-based similarity matching is highly ambiguous at vocabulary scale. These results support the claim that AlienLM reduces human-readable exposure at the API boundary under practical observer access patterns.

%% file: sections/12_appendix_safety.tex
  \section{Safety and Alignment Evaluation}
  \label{app:safety}

  This appendix reports additional details for the safety/alignment results summarized in the Impact Statement.
  Our goal is to assess whether API-only adaptation on alienized data preserves the model's refusal and safety behaviors.

  \paragraph{Benchmarks.}
  We evaluate eight public safety benchmarks covering harmful instruction following, toxicity, jailbreak robustness, and trustworthiness:
  WildGuardTest (WildG), HarmBench, ToxiGen, XSTest, WildJailbreak (benign/harmful splits; WildJ-b/WildJ-h), DAN, and TrustLLM.
  All metrics are normalized to a 0--100 scale where higher is safer (100 indicates best safety performance for that benchmark).

  \paragraph{Models and settings.}
  We compare each backbone's \textbf{Oracle} (base model without alienization) and its \textbf{AlienLM} variant (AAT with $\rho{=}1$).
  All AlienLM variants are adapted using the same AAT protocol described in Section~\ref{subsec:adaptation}.

  \paragraph{Findings.}
  As shown in Table~\ref{tab:safety}, AlienLM tends to reduce average safety scores across all models, with the largest drops appearing on jailbreak-focused benchmarks (e.g., WildJ-h, DAN) and trustworthiness metrics.
  This suggests that adapting models to operate on alienized inputs can alter refusal and safety behaviors even when training data are benign and utility-oriented.

  \paragraph{Interpretation and limitation.}
  AlienLM optimizes for utility recovery under alienized text, not for preserving alignment. Therefore, safety regressions should be interpreted as a practical limitation of the current approach rather than an intended outcome.
  This is consistent with prior evidence that fine-tuning can unintentionally shift safety behavior.
  Improving safety retention under AAT is an important direction for future work.

  \paragraph{Mitigation directions (non-exhaustive).}
  Potential mitigations include (i) safety-aware adaptation objectives (e.g., mixing a small set of safety preference data during AAT),
  (ii) post-hoc safety re-alignment on alienized safety prompts, and (iii) lightweight client-side filtering as a complementary layer.
  We leave a systematic study to future work.

%% file: sections/11_appendix_qualitative.tex
\section{Qualitative Examples of Alien Language}
\label{app:qual}

This appendix provides qualitative examples complementing Section~\ref{subsec:robustness} and Section~\ref{subsec:adaptation}. The goal is to illustrate two properties of \alienlm:
(i) \emph{human opacity} at the API boundary (alien text is difficult to interpret without the translator),
and (ii) \emph{model-side consistency} (the same alien tokens repeat consistently, enabling learnability under AAT).

\paragraph{Rendering notes.}
Tokenizers used by modern LLMs (e.g., BPE/byte-fallback variants) may include tokens corresponding to non-printable byte sequences. For readability, we omit such tokens in the examples when they do not render cleanly. When a token string exists but cannot be rendered in \LaTeX{} due to Unicode limitations, we display it as \texttt{<<UNICODE>>}. These display choices do not affect the underlying token-ID mapping, which remains lossless.

\subsection{Reasoning Example (GSM8K)}
\label{app:qual-gsm8k}

\paragraph{Observation.}
The alien text exhibits several notable patterns. First, numbers are mapped to numbers of similar scale (e.g., ``16'' $\to$ ``116'', ``18'' $\to$ ``181'', ``2'' $\to$ ``212''), likely because our embedding-based bijection pairs semantically related tokens. This partial numerical consistency may explain why AlienLM substantially outperforms random bijection on math reasoning tasks (Table~\ref{tab:main}). Second, semantic tokens are replaced with unrelated terms (e.g., ``eggs'' $\to$ ``jars'', ``ducks'' $\to$ ``Beetle''), obscuring content while the model internally maintains consistent mappings. Third, repeated tokens appear consistently across the alien question and answer, reflecting the deterministic token-ID relabeling and enabling the model to learn stable input-output associations during AAT.

\subsection{Code Example (MBPP)}
\label{app:qual-code}

  \paragraph{Observation.}
  The alienized code exhibits several patterns distinct from natural language examples. First, variable names are mapped consistently across occurrences (e.g., \texttt{s\_list} $\to$ \texttt{productList} appears multiple times), enabling the model to track variable references during AAT. Second, semantically related tokens are sometimes mapped to surface-similar terms (e.g., \texttt{vowels} $\to$ \texttt{towels}), reflecting the embedding-based bijection's tendency to pair related concepts. Third, a large fraction of tokens are mapped at the byte level, producing unreadable identifiers (e.g., \texttt{<<UNICODE>>}, \texttt{IOExceptionZa}). This preserves token-level consistency for the model while rendering the code nearly impossible for humans to interpret, even for those familiar with programming syntax.

\newpage

\begin{table*}
  \centering
  \caption{Ablations on LLaMA~3~8B (accuracy, \%). \textsc{Average} is the unweighted mean over seven benchmarks. ${}^{\dagger}$ uses proxy head $e_P$ under the black-box constraint. \cradd{${}^{\ddagger}$ uses full white-box access (gradient-based embedding alignment); included as an upper-bound reference, not a deployable baseline.}}
  \label{tab:ablation}
  \adjustbox{width=\textwidth}{
    \begin{tabular}{llcccccccc}
      \toprule
      \textbf{Methods}& \textbf{Components}& MMLU & ARC-E & ARC-C & HellaS & WinoG & TQA & GSM8K & Average \\
      \midrule
      \multirow{3}{*}{\makecell{LLaMA~3 \\ 8B}}
      & Oracle & 67.32 & 84.13 & 59.39 & 57.07 & 74.35 & 35.25 & 75.89 & 64.77 \\
      & SFT & 63.74 & 80.56 & 53.67 & 53.70 & 71.74 & 37.58 & 76.12 & 62.44 \\
      & \cradd{SentinelLM (white-box)${}^{\ddagger}$} & \cradd{63.67} & \cradd{80.77} & \cradd{53.50} & \cradd{54.16} & \cradd{72.61} & \cradd{37.58} & \cradd{74.45} & \cradd{62.39} \\
      \midrule
      \multirow{4}{*}{AlienLM}
      & $e_P$ LM Head${}^{\dagger}$ & 49.42 & 72.14 & \textbf{44.28} & 47.86 & 61.48 & 35.01 & 63.08 & 53.32 \\
      & $e_{\text{tgt}}$ LM Head & \textbf{51.60} & \textbf{73.73} & 44.20 & \textbf{48.38} & \textbf{65.11} & \textbf{36.96} & \textbf{65.50} & \textbf{55.07} \\
      & $e_{\text{tgt}}$ Embeddings & 50.82 & 68.64 & 43.67 & 47.98 & 64.01 & 36.47 & 64.14 & 53.68 \\
      & Random $\mathcal{V}$ & 29.92 & 46.34 & 27.56 & 38.47 & 55.09 & 30.23 & 31.08 & 36.96 \\
      \bottomrule
    \end{tabular}
  }
\end{table*}

\begin{table*}
  \centering
  \caption{Domain-specific fine-tuning on general benchmarks (LLaMA~3~8B).
  \textsc{Average} over MMLU, ARC-E, ARC-C, HellaSwag, WinoGrande, TruthfulQA, GSM8K.}
  \label{tab:finetune-general}
  \adjustbox{width=\textwidth}{
    \begin{tabular}{lcc|ccccccc|c}
      \toprule
      \textbf{Method} & \textbf{Tokenizer} & \textbf{Data} &
      \textbf{MMLU} & \textbf{ARC-E} & \textbf{ARC-C} & \textbf{HellaS} &
      \textbf{WinoG} & \textbf{TQA} & \textbf{GSM8K} & \textbf{Average} \\
      \midrule
      full           & O    & O    & 45.59 & 70.58 & 42.41 & 47.32 & 61.25 & 31.21 & 55.50 & 50.55\\
      - tokenizer    & X    & O    & 46.13 & 70.71 & 41.89 & 47.62 & 58.96 & 31.82 & 56.94 & 50.58\\
      - data         & O    & X    & 47.50 & 71.68 & 42.06 & 47.86 & 61.56 & 31.95 & 41.70 & 49.19\\
      - tok \& data  & X    & X    & 47.26 & 72.39 & 43.26 & 47.72 & 59.27 & 33.17 & 41.02 & 49.16\\
      code/math only & only & only & 28.68 & 57.70 & 35.84 & 40.71 & 56.20 & 34.88 & 55.04 & 44.15\\
      + data         & O    & +150k& 45.18 & 71.55 & 43.09 & 48.15 & 62.75 & 32.80 & 60.60 & 52.02\\
      \bottomrule
    \end{tabular}
  }
\end{table*}

\begin{table*}
  \centering
  \caption{Seed diversity results.}
  \label{tab:seed-diversity}
  \label{tab:keydiv}

  \begin{adjustbox}{width=\textwidth}
    \begin{tabular}{llcc|ccccccc}
      \toprule
      \textbf{Models} & \textbf{Method} & \textbf{Average} & \textbf{Ratio} &
      \textbf{MMLU} & \textbf{ARC-E} & \textbf{ARC-C} & \textbf{HellaSwag} &
      \textbf{WinoG} & \textbf{TQA} & \textbf{GSM8K} \\
      \midrule
      \multirow{12}{*}{LLaMA~3~8B}
      & Oracle          & 64.77 & -    & 67.32 & 84.13 & 59.39 & 57.07 & 74.35 & 35.25 & 75.89 \\
      & Random            & 36.96 & 57.06 & 29.92 & 46.34 & 27.56 & 38.47 & 55.09 & 30.23 & 31.08 \\
      & AlienLM           & 52.92 & 81.70 & 46.56 & 72.14 & 44.28 & 47.86 & 61.48 & 35.01 & 63.08 \\
      \cmidrule(lr){2-11}
      & \multicolumn{10}{l}{\textit{bucketed pairing}} \\
      & \hspace{1em} seed=42  & 50.98 & 78.71 & 45.59 & 67.47 & 42.49 & 46.80 & 60.69 & 34.15 & 59.67 \\
      & \hspace{1em} seed=43  & 51.16 & 78.98 & 44.82 & 67.93 & 42.75 & 46.15 & 61.01 & 33.41 & 62.02 \\
      & \hspace{1em} seed=44  & 50.45 & 77.89 & 44.61 & 67.80 & 42.15 & 46.68 & 60.22 & 32.93 & 58.76 \\
      & \hspace{1em} seed=45  & 51.57 & 79.61 & 47.24 & 66.88 & 42.92 & 46.78 & 61.64 & 34.39 & 61.11 \\
      & \hspace{1em} seed=46  & 50.62 & 78.15 & 46.07 & 67.00 & 43.86 & 46.57 & 58.64 & 34.27 & 57.92 \\
      \cmidrule(lr){2-11}
      & \hspace{1em} \textbf{Mean} & 51.28& 79.17& 45.82& 68.20& 43.08& 46.81& 60.61& 34.03& 60.43\\
      & \hspace{1em} \textbf{Std}  & 0.89& 1.38& 1.01& 1.97& 0.82& 0.57& 1.10& 0.74& 1.98\\
      \bottomrule
    \end{tabular}
  \end{adjustbox}
\end{table*}

\begin{table*}
  \centering
  \caption{Safety/alignment results comparing Oracle vs.\ AlienLM across target models and benchmarks (0--100, higher is safer). WildG = WildGuardTest; WildJ-b/h = WildJailbreak benign/harmful; DAN = Do-Anything-Now.}
  \label{tab:safety}
  \adjustbox{width=\textwidth}{
    \begin{tabular}{llccccccccc}
      \toprule
      \textbf{Model} & \textbf{Method} & \textbf{WildG} & \textbf{HarmBench} & \textbf{ToxiGen} & \textbf{XSTest} & \textbf{WildJ-b} & \textbf{WildJ-h} & \textbf{DAN} & \textbf{TrustLLM} & \textbf{AVG} \\
      \midrule
      \multirow{2}{*}{LLaMA~3~8B} & Oracle & 5.21 & 84.69 & 100.00 & 96.89 & 8.40 & 80.50 & 98.70 & 89.50 & 70.49 \\
      & AlienLM & 28.84 & 55.00 & 99.29 & 75.33 & 0.80 & 7.35 & 20.67 & 51.25 & 42.32 \\
      \midrule
      \multirow{2}{*}{Qwen~2.5-7B} & Oracle & 15.09 & 83.44 & 100.00 & 92.22 & 0.40 & 13.85 & 64.33 & 68.25 & 54.70 \\
      & AlienLM & 31.11 & 43.44 & 95.21 & 72.67 & 0.80 & 2.55 & 29.00 & 50.25 & 40.63 \\
      \midrule
      \multirow{2}{*}{Qwen~2.5-14B} & Oracle & 9.21 & 93.44 & 100.00 & 94.00 & 0.80 & 25.10 & 23.30 & 19.50 & 45.67 \\
      & AlienLM & 34.45 & 59.06 & 100.00 & 76.44 & 0.40 & 1.75 & 11.00 & 48.25 & 41.42 \\
      \midrule
      \multirow{2}{*}{Gemma~2 9B} & Oracle & 10.41 & 92.81 & 100.00 & 90.00 & 1.60 & 42.35 & 32.33 & 80.25 & 56.22 \\
      & AlienLM & 34.98 & 44.06 & 99.71 & 75.11 & 1.60 & 3.30 & 8.67 & 47.00 & 39.30 \\
      \bottomrule
    \end{tabular}
  }
\end{table*}

\tcbset{
  wrap/.style={
    title=Natural vs. Alien (illustrative),
    colback=gray!4,
    colframe=black!15,
    colbacktitle=gray!15,
    coltitle=black!80,
    breakable,
    enhanced jigsaw,
    boxsep=2.2mm,
    left=0.5mm, right=0.5mm, top=0.5mm, bottom=0.5mm,
  },
  alien/.style={
    colback=mygreenlight,
    colframe=mygreendark,
    title=Alien Language,
    fonttitle=\ttfamily,
    fontupper=\ttfamily,
    breakable,
    enhanced jigsaw,
    width=\linewidth
  },
  natural/.style={
    colback=blue!8,
    colframe=blue!30,
    title=Natural Language,
    fonttitle=\ttfamily,
    fontupper=\ttfamily,
    breakable,
    enhanced jigsaw,
    width=\linewidth
  }
}

\begin{figure}
  \centering
  \caption{Qualitative example from GSM8K. The alienized output preserves numerical structure (e.g., ``16'' $\to$ ``116'', ``18'' $\to$ ``181'') while rendering the text unreadable to humans.}
  \label{fig:qual-gsm8k}
  \begin{tcolorbox}[
      title=natural,
      width=\columnwidth,
      ]
      \textbf{Q:} Janet’s ducks lay 16 eggs per day. She eats three for breakfast and bakes muffins with four. She sells the remainder for \$2 per egg. How much does she make per day?

    \textbf{A:} She has $16-3-4=9$ eggs left. $9\times 2=18$. The answer is 18.
  \end{tcolorbox}
  \vspace{2mm}
  \begin{tcolorbox}[alien]
  \textbf{Q:} Feb Beetle———————————————— ck anlayış 116 gs perpetual Everyday, Cats five withreak ning but Ib tweaking puff insulting with girlfriends Everyday for six, dell aremainder zatím a Bers` supermarketsdaily with 212 perpetual freshwaterduckegg, HOW hodin jars doesnt Everyday zatím a Bers` supermarkets!
    
    \textbf{A:} Jan overlays116 gs perpetual Everyday,Cats 433 but Ib tweaking 344, Sox 116 gs right,dell 998 gs with 212 jars , 998 data 212 181, A answered 181;
  \end{tcolorbox}
\end{figure}

\begin{figure}
  \centering
  \caption{Qualitative example from MBPP (code generation). Variable names are consistently mapped (e.g., \texttt{s\_list} $\to$ \texttt{productList}), while byte-level tokens produce unreadable identifiers.}
  \label{fig:qual-mbpp}
  \begin{tcolorbox}[
      title=natural,
      width=\columnwidth,
    ]
    \textbf{Q:} Write a Python function to reverse only the vowels of a given string.

    \textbf{A:}
\begin{lstlisting}[language=Python,breaklines=true,breakatwhitespace=true,columns=fullflexible]
def reverse_vowels(s):
    vowels = 'aeiouAEIOU'
    s_list = list(s)
    left, right = 0, len(s) - 1
    while left < right:
        if s_list[left] not in vowels:
            left += 1
        elif s_list[right] not in vowels:
            right -= 1
        else:
            s_list[left], s_list[right] = s_list[right], s_list[left]
            left += 1
            right -= 1
    return ''.join(s_list)
\end{lstlisting}
  \end{tcolorbox}
  \vspace{2mm}
  \begin{tcolorbox}[alien]
  \sloppy
    \textbf{Q:} Read the/python functionality reverse a towels the stringByAppendingString,

    \textbf{A:} getopt(t productList]tyleAttrr everse\_w blowjob Michaels(t": towels "<<UNICODE>> iNdEx<<UNICODE>>IOExceptionZa" helps productList (t] right. left 100. düzenlenen(t] \/ 211 right left?\; ...
    \end{tcolorbox}
  \end{figure}

%% file: iclr2026_conference.bib
@inproceedings{gilad-bachrach2016cryptonets,
  author    = {Ran Gilad{-}Bachrach and Nathan Dowlin and Kim Laine and Kristin Lauter and Michael Naehrig and John Wernsing},
  title     = {CryptoNets: Applying Neural Networks to Encrypted Data with High Throughput and Accuracy},
  booktitle = {Proceedings of the 33rd International Conference on Machine Learning (ICML)},
  year      = {2016},
  pages     = {201--210},
  url       = {https://proceedings.mlr.press/v48/gilad-bachrach16.html}
}

@inproceedings{juvekar2018gazelle,
  author    = {Chiraag Juvekar and Vinod Vaikuntanathan and Anantha Chandrakasan},
  title     = {G{A}{Z}{E}{L}{L}{E}: A Low Latency Framework for Secure Neural Network Inference},
  booktitle = {27th USENIX Security Symposium},
  year      = {2018},
  pages     = {1651--1669},
  url       = {https://www.usenix.org/conference/usenixsecurity18/presentation/juvekar}
}

@inproceedings{mishra2020delphi,
  author    = {Pratyush Mishra and Rishabh Poddar and Sameer Wagh and Shafi Goldwasser and Raluca Ada Popa and Joseph E. Gonzalez and Dawn Song},
  title     = {Delphi: A Cryptographic Inference Service for Neural Networks},
  booktitle = {29th USENIX Security Symposium},
  year      = {2020},
  pages     = {2505--2522},
  url       = {https://www.usenix.org/conference/usenixsecurity20/presentation/mishra}
}

@inproceedings{li2021dp,
  author    = {Xuechen Li and Florian Tramer and Percy Liang and Tatsunori Hashimoto},
  title     = {Large Language Models Can Be Strong Differentially Private Learners},
  booktitle = {International Conference on Learning Representations (ICLR)},
  year      = {2022},
  note      = {Oral},
  url       = {https://iclr.cc/virtual/2022/oral/6895}
}

@article{yao2024fedllmsurvey,
  author    = {Yuhang Yao and Jianyi Zhang and Junda Wu and Chengkai Huang and Yu Xia and Tong Yu and Ruiyi Zhang and Sungchul Kim and Ryan Rossi and Ang Li and Lina Yao and Julian McAuley and Yiran Chen and Carlee Joe-Wong},
  title     = {Federated Large Language Models: Current Progress and Future Directions},
  journal   = {arXiv preprint arXiv:2409.15723},
  year      = {2024},
  doi       = {10.48550/arXiv.2409.15723},
  note      = {Survey},
  url       = {https://arxiv.org/abs/2409.15723}
}

@article{fedllmbench2024,
  author    = {Rui Ye and Rui Ge and Xinyu Zhu and Jingyi Chai and Yaxin Du and Yang Liu and Yanfeng Wang and Siheng Chen},
  title     = {FedLLM-Bench: Realistic Benchmarks for Federated Learning of Large Language Models},
  journal   = {arXiv preprint arXiv:2406.04845},
  year      = {2024},
  doi       = {10.48550/arXiv.2406.04845},
  note      = {Benchmark},
  url       = {https://arxiv.org/abs/2406.04845}
}

@inproceedings{kornblith2019cka,
  author    = {Simon Kornblith and Mohammad Norouzi and Honglak Lee and Geoffrey Hinton},
  title     = {Similarity of Neural Network Representations Revisited},
  booktitle = {ICML},
  year      = {2019}
}

@inproceedings{bansal2021stitching,
  author    = {Yash Sharma Bansal and others},
  title     = {Revisiting Model Stitching to Compare Neural Representations},
  booktitle = {NeurIPS},
  year      = {2021}
}

@inproceedings{
langplasticity,
title={Improving Language Plasticity via Pretraining with Active Forgetting},
author={Yihong Chen and Kelly Marchisio and Roberta Raileanu and David Ifeoluwa Adelani and Pontus Stenetorp and Sebastian Riedel and Mikel Artetxe},
booktitle={Thirty-seventh Conference on Neural Information Processing Systems},
year={2023},
url={https://openreview.net/forum?id=jvEbQBxd8X}
}

@inproceedings{
magpie,
title={Magpie: Alignment Data Synthesis from Scratch by Prompting Aligned {LLM}s with Nothing},
author={Zhangchen Xu and Fengqing Jiang and Luyao Niu and Yuntian Deng and Radha Poovendran and Yejin Choi and Bill Yuchen Lin},
booktitle={The Thirteenth International Conference on Learning Representations},
year={2025},
url={https://openreview.net/forum?id=Pnk7vMbznK}
}

@misc{gemmateam2024gemma2improvingopen,
      title={Gemma 2: Improving Open Language Models at a Practical Size}, 
      author={Gemma Team and Morgane Riviere and Shreya Pathak and Pier Giuseppe Sessa and Cassidy Hardin and Surya Bhupatiraju and Léonard Hussenot and Thomas Mesnard and Bobak Shahriari and Alexandre Ramé and Johan Ferret and Peter Liu and Pouya Tafti and Abe Friesen and Michelle Casbon and Sabela Ramos and Ravin Kumar and Charline Le Lan and Sammy Jerome and Anton Tsitsulin and Nino Vieillard and Piotr Stanczyk and Sertan Girgin and Nikola Momchev and Matt Hoffman and Shantanu Thakoor and Jean-Bastien Grill and Behnam Neyshabur and Olivier Bachem and Alanna Walton and Aliaksei Severyn and Alicia Parrish and Aliya Ahmad and Allen Hutchison and Alvin Abdagic and Amanda Carl and Amy Shen and Andy Brock and Andy Coenen and Anthony Laforge and Antonia Paterson and Ben Bastian and Bilal Piot and Bo Wu and Brandon Royal and Charlie Chen and Chintu Kumar and Chris Perry and Chris Welty and Christopher A. Choquette-Choo and Danila Sinopalnikov and David Weinberger and Dimple Vijaykumar and Dominika Rogozińska and Dustin Herbison and Elisa Bandy and Emma Wang and Eric Noland and Erica Moreira and Evan Senter and Evgenii Eltyshev and Francesco Visin and Gabriel Rasskin and Gary Wei and Glenn Cameron and Gus Martins and Hadi Hashemi and Hanna Klimczak-Plucińska and Harleen Batra and Harsh Dhand and Ivan Nardini and Jacinda Mein and Jack Zhou and James Svensson and Jeff Stanway and Jetha Chan and Jin Peng Zhou and Joana Carrasqueira and Joana Iljazi and Jocelyn Becker and Joe Fernandez and Joost van Amersfoort and Josh Gordon and Josh Lipschultz and Josh Newlan and Ju-yeong Ji and Kareem Mohamed and Kartikeya Badola and Kat Black and Katie Millican and Keelin McDonell and Kelvin Nguyen and Kiranbir Sodhia and Kish Greene and Lars Lowe Sjoesund and Lauren Usui and Laurent Sifre and Lena Heuermann and Leticia Lago and Lilly McNealus and Livio Baldini Soares and Logan Kilpatrick and Lucas Dixon and Luciano Martins and Machel Reid and Manvinder Singh and Mark Iverson and Martin Görner and Mat Velloso and Mateo Wirth and Matt Davidow and Matt Miller and Matthew Rahtz and Matthew Watson and Meg Risdal and Mehran Kazemi and Michael Moynihan and Ming Zhang and Minsuk Kahng and Minwoo Park and Mofi Rahman and Mohit Khatwani and Natalie Dao and Nenshad Bardoliwalla and Nesh Devanathan and Neta Dumai and Nilay Chauhan and Oscar Wahltinez and Pankil Botarda and Parker Barnes and Paul Barham and Paul Michel and Pengchong Jin and Petko Georgiev and Phil Culliton and Pradeep Kuppala and Ramona Comanescu and Ramona Merhej and Reena Jana and Reza Ardeshir Rokni and Rishabh Agarwal and Ryan Mullins and Samaneh Saadat and Sara Mc Carthy and Sarah Cogan and Sarah Perrin and Sébastien M. R. Arnold and Sebastian Krause and Shengyang Dai and Shruti Garg and Shruti Sheth and Sue Ronstrom and Susan Chan and Timothy Jordan and Ting Yu and Tom Eccles and Tom Hennigan and Tomas Kocisky and Tulsee Doshi and Vihan Jain and Vikas Yadav and Vilobh Meshram and Vishal Dharmadhikari and Warren Barkley and Wei Wei and Wenming Ye and Woohyun Han and Woosuk Kwon and Xiang Xu and Zhe Shen and Zhitao Gong and Zichuan Wei and Victor Cotruta and Phoebe Kirk and Anand Rao and Minh Giang and Ludovic Peran and Tris Warkentin and Eli Collins and Joelle Barral and Zoubin Ghahramani and Raia Hadsell and D. Sculley and Jeanine Banks and Anca Dragan and Slav Petrov and Oriol Vinyals and Jeff Dean and Demis Hassabis and Koray Kavukcuoglu and Clement Farabet and Elena Buchatskaya and Sebastian Borgeaud and Noah Fiedel and Armand Joulin and Kathleen Kenealy and Robert Dadashi and Alek Andreev},
      year={2024},
      eprint={2408.00118},
      doi={10.48550/arXiv.2408.00118},
      archivePrefix={arXiv},
      primaryClass={cs.CL},
      url={https://arxiv.org/abs/2408.00118}, 
}

@article{mmlu,
  title={Measuring Massive Multitask Language Understanding},
  author={Dan Hendrycks and Collin Burns and Steven Basart and Andy Zou and Mantas Mazeika and Dawn Song and Jacob Steinhardt},
  journal={Proceedings of the International Conference on Learning Representations (ICLR)},
  year={2021}
}

@misc{arc,
      title={Think you have Solved Question Answering? Try ARC, the AI2 Reasoning Challenge}, 
      author={Peter Clark and Isaac Cowhey and Oren Etzioni and Tushar Khot and Ashish Sabharwal and Carissa Schoenick and Oyvind Tafjord},
      year={2018},
      eprint={1803.05457},
      archivePrefix={arXiv},
      primaryClass={cs.AI},
      doi={10.48550/arXiv.1803.05457},
      url={https://arxiv.org/abs/1803.05457}, 
}

@article{winogrande,
author = {Sakaguchi, Keisuke and Bras, Ronan Le and Bhagavatula, Chandra and Choi, Yejin},
title = {WinoGrande: an adversarial winograd schema challenge at scale},
year = {2021},
issue_date = {September 2021},
publisher = {Association for Computing Machinery},
address = {New York, NY, USA},
volume = {64},
number = {9},
issn = {0001-0782},
url = {https://doi.org/10.1145/3474381},
doi = {10.1145/3474381},
journal = {Commun. ACM},
month = aug,
pages = {99–106},
numpages = {8}
}

@inproceedings{truthfulqa,
    title = "{T}ruthful{QA}: Measuring How Models Mimic Human Falsehoods",
    author = "Lin, Stephanie  and
      Hilton, Jacob  and
      Evans, Owain",
    editor = "Muresan, Smaranda  and
      Nakov, Preslav  and
      Villavicencio, Aline",
    booktitle = "Proceedings of the 60th Annual Meeting of the Association for Computational Linguistics (Volume 1: Long Papers)",
    month = may,
    year = "2022",
    address = "Dublin, Ireland",
    publisher = "Association for Computational Linguistics",
    url = "https://aclanthology.org/2022.acl-long.229/",
    doi = "10.18653/v1/2022.acl-long.229",
    pages = "3214--3252"
}

@misc{gsm8k,
      title={Training Verifiers to Solve Math Word Problems}, 
      author={Karl Cobbe and Vineet Kosaraju and Mohammad Bavarian and Mark Chen and Heewoo Jun and Lukasz Kaiser and Matthias Plappert and Jerry Tworek and Jacob Hilton and Reiichiro Nakano and Christopher Hesse and John Schulman},
      year={2021},
      eprint={2110.14168},
      archivePrefix={arXiv},
      primaryClass={cs.LG},
      doi={10.48550/arXiv.2110.14168},
      url={https://arxiv.org/abs/2110.14168}, 
}

@inproceedings{hellaswag,
    title={HellaSwag: Can a Machine Really Finish Your Sentence?},
    author={Zellers, Rowan and Holtzman, Ari and Bisk, Yonatan and Farhadi, Ali and Choi, Yejin},
    booktitle ={Proceedings of the 57th Annual Meeting of the Association for Computational Linguistics},
    year={2019}
}

@inproceedings{sentinellm,
author = {Mishra, Abhijit and Li, Mingda and Deo, Soham},
title = {SentinelLMs: encrypted input adaptation and fine-tuning of language models for private and secure inference},
booktitle = {Proceedings of the AAAI Conference on Artificial Intelligence},
year = {2024},
isbn = {978-1-57735-887-9},
publisher = {AAAI Press},
url = {https://doi.org/10.1609/aaai.v38i19.30136},
doi = {10.1609/aaai.v38i19.30136},
articleno = {2388},
numpages = {9},
series = {AAAI'24/IAAI'24/EAAI'24}
}

@misc{slimOrca,
  title = {SlimOrca: An Open Dataset of GPT-4 Augmented FLAN Reasoning Traces, with Verification},
  author = {Wing Lian and Guan Wang and Bleys Goodson and Eugene Pentland and Austin Cook and Chanvichet Vong and "Teknium"},
  year = {2023},
  publisher = {HuggingFace},
  url = {https://https://huggingface.co/Open-Orca/SlimOrca}
}

@inproceedings{li2022dpstrong,
  title={Large Language Models Can Be Strong Differentially Private Learners},
  author={Li, Xuechen and Tram{\`e}r, Florian and Liang, Percy and Hashimoto, Tatsunori},
  booktitle={ICLR},
  year={2022}
}

@article{Dubey2024Llama3,
  author = {A. Dubey and A. Jauhri and A. Pandey and A. Kadian and A. Al-Dahle and A. Letman and A. Mathur and A. Schelten and A. Yang and A. Fan and others},
  title = {The Llama 3 Herd of Models},
  journal = {arXiv preprint arXiv:2407.21783},
  year = {2024},
  doi = {10.48550/arXiv.2407.21783},
  url = {https://arxiv.org/abs/2407.21783}
}

@article{Yang2024Qwen2_5,
  author       = {An Yang and others},
  title        = {Qwen\,2.5 Technical Report},
  journal      = {arXiv preprint arXiv:2412.15115},
  year         = {2024},
  doi          = {10.48550/arXiv.2412.15115},
  url          = {https://arxiv.org/abs/2412.15115}
}

@inproceedings{
sparsefeature,
title={Sparse Autoencoders Find Highly Interpretable Features in Language Models},
author={Robert Huben and Hoagy Cunningham and Logan Riggs Smith and Aidan Ewart and Lee Sharkey},
booktitle={The Twelfth International Conference on Learning Representations},
year={2024},
url={https://openreview.net/forum?id=F76bwRSLeK}
}

@inproceedings{languagefeature,
    title = "Unveiling Language-Specific Features in Large Language Models via Sparse Autoencoders",
    author = "Deng, Boyi  and
      Wan, Yu  and
      Yang, Baosong  and
      Zhang, Yidan  and
      Feng, Fuli",
    editor = "Che, Wanxiang  and
      Nabende, Joyce  and
      Shutova, Ekaterina  and
      Pilehvar, Mohammad Taher",
    booktitle = "Proceedings of the 63rd Annual Meeting of the Association for Computational Linguistics (Volume 1: Long Papers)",
    month = jul,
    year = "2025",
    address = "Vienna, Austria",
    publisher = "Association for Computational Linguistics",
    url = "https://aclanthology.org/2025.acl-long.229/",
    doi = "10.18653/v1/2025.acl-long.229",
    pages = "4563--4608",
    ISBN = "979-8-89176-251-0",
}

@inproceedings{dpsgd,
author = {Abadi, Martin and Chu, Andy and Goodfellow, Ian and McMahan, H. Brendan and Mironov, Ilya and Talwar, Kunal and Zhang, Li},
title = {Deep Learning with Differential Privacy},
year = {2016},
isbn = {9781450341394},
publisher = {Association for Computing Machinery},
address = {New York, NY, USA},
url = {https://doi.org/10.1145/2976749.2978318},
doi = {10.1145/2976749.2978318},
booktitle = {Proceedings of the 2016 ACM SIGSAC Conference on Computer and Communications Security},
pages = {308–318},
numpages = {11},
location = {Vienna, Austria},
series = {CCS '16}
}

@inproceedings{emojiprompt,
    title = "{E}moji{P}rompt: Generative Prompt Obfuscation for Privacy-Preserving Communication with Cloud-based {LLM}s",
    author = "Lin, Sam  and
      Hua, Wenyue  and
      Wang, Zhenting  and
      Jin, Mingyu  and
      Fan, Lizhou  and
      Zhang, Yongfeng",
    editor = "Chiruzzo, Luis  and
      Ritter, Alan  and
      Wang, Lu",
    booktitle = "Proceedings of the 2025 Conference of the Nations of the Americas Chapter of the Association for Computational Linguistics: Human Language Technologies (Volume 1: Long Papers)",
    month = apr,
    year = "2025",
    address = "Albuquerque, New Mexico",
    publisher = "Association for Computational Linguistics",
    url = "https://aclanthology.org/2025.naacl-long.614/",
    doi = "10.18653/v1/2025.naacl-long.614",
    pages = "12342--12361",
    ISBN = "979-8-89176-189-6",
}

@inproceedings{
dettmers2022bit,
title={8-bit Optimizers via Block-wise Quantization},
author={Tim Dettmers and Mike Lewis and Sam Shleifer and Luke Zettlemoyer},
booktitle={International Conference on Learning Representations},
year={2022},
url={https://openreview.net/forum?id=shpkpVXzo3h}
}

@article{johnson2019billion,
  title={Billion-scale similarity search with {GPUs}},
  author={Johnson, Jeff and Douze, Matthijs and J{\'e}gou, Herv{\'e}},
  journal={IEEE Transactions on Big Data},
  volume={7},
  number={3},
  pages={535--547},
  year={2019},
  publisher={IEEE}
}

@misc{mbpp,
      title={Program Synthesis with Large Language Models}, 
      author={Jacob Austin and Augustus Odena and Maxwell Nye and Maarten Bosma and Henryk Michalewski and David Dohan and Ellen Jiang and Carrie Cai and Michael Terry and Quoc Le and Charles Sutton},
      year={2021},
      eprint={2108.07732},
      archivePrefix={arXiv},
      primaryClass={cs.PL},
      doi={10.48550/arXiv.2108.07732},
      url={https://arxiv.org/abs/2108.07732}, 
}

@misc{humaneval,
  title={Evaluating Large Language Models Trained on Code},
  author={Mark Chen and Jerry Tworek and Heewoo Jun and Qiming Yuan and Henrique Ponde de Oliveira Pinto and Jared Kaplan and Harri Edwards and Yuri Burda and Nicholas Joseph and Greg Brockman and Alex Ray and Raul Puri and Gretchen Krueger and Michael Petrov and Heidy Khlaaf and Girish Sastry and Pamela Mishkin and Brooke Chan and Scott Gray and Nick Ryder and Mikhail Pavlov and Alethea Power and Lukasz Kaiser and Mohammad Bavarian and Clemens Winter and Philippe Tillet and Felipe Petroski Such and Dave Cummings and Matthias Plappert and Fotios Chantzis and Elizabeth Barnes and Ariel Herbert-Voss and William Hebgen Guss and Alex Nichol and Alex Paino and Nikolas Tezak and Jie Tang and Igor Babuschkin and Suchir Balaji and Shantanu Jain and William Saunders and Christopher Hesse and Andrew N. Carr and Jan Leike and Josh Achiam and Vedant Misra and Evan Morikawa and Alec Radford and Matthew Knight and Miles Brundage and Mira Murati and Katie Mayer and Peter Welinder and Bob McGrew and Dario Amodei and Sam McCandlish and Ilya Sutskever and Wojciech Zaremba},
  year={2021},
  eprint={2107.03374},
  archivePrefix={arXiv},
  primaryClass={cs.LG},
  doi={10.48550/arXiv.2107.03374},
  url={https://arxiv.org/abs/2107.03374}
}

@inproceedings{qi2024finetuning,
    title = "Fine-tuning Aligned Language Models Compromises Safety, Even When Users Do Not Intend To!",
    author = "Qi, Xiangyu and Zeng, Yi and Xie, Tinghao and Chen, Pin-Yu and Jia, Ruoxi and Mittal, Prateek and Henderson, Peter",
    booktitle = "The Twelfth International Conference on Learning Representations",
    year = "2024",
    url = "https://openreview.net/forum?id=hTEGyKf0dZ"
}

@inproceedings{halawi2024covert,
    title = "Covert Malicious Finetuning: Challenges in Safeguarding {LLM} Adaptation",
    author = "Halawi, Danny and Wei, Alexander and Wallace, Eric and Steinhardt, Jacob",
    booktitle = "Forty-first International Conference on Machine Learning",
    year = "2024",
    url = "https://openreview.net/forum?id=Bb5f0hFBbq"
}

@article{thor2024,
  title={{THOR}: Secure Transformer Inference with Homomorphic Encryption},
  author={Moon, Jungho and others},
  journal={IACR Cryptology ePrint Archive},
  year={2024},
  url={https://eprint.iacr.org/2024/1881}
}

@inproceedings{secformer2024,
  title     = {{SecFormer}: Fast and Accurate Privacy-Preserving Inference for Transformer Models via {SMPC}},
  author    = {Jinglong Luo and Yehong Zhang and Zhuo Zhang and Jiaqi Zhang and Xin Mu and Hui Wang and Yue Yu and Zenglin Xu},
  booktitle = {Findings of the Association for Computational Linguistics: ACL 2024},
  year      = {2024},
  pages     = {13333--13348},
  doi       = {10.18653/v1/2024.findings-acl.790},
  url       = {https://aclanthology.org/2024.findings-acl.790/}
}

@article{teeconfidential2024,
  title={Confidential LLM Inference: Performance and Cost Across CPU and GPU TEEs},
  author={Marcin Chrapek and Marcin Copik and Etienne Mettaz and Torsten Hoefler},
  journal={arXiv preprint arXiv:2509.18886},
  year={2025},
  doi={10.48550/arXiv.2509.18886},
  url={https://arxiv.org/abs/2509.18886},
  note={Preprint}
}

@article{inferdpt2025,
  title={{InferDPT}: Privacy-Preserving Inference for Black-box Large Language Models},
  author={Tong, Meng and others},
  journal={IEEE Transactions on Dependable and Secure Computing},
  year={2025}
}

@inproceedings{papillon2025,
  title={{PAPILLON}: Privacy Preservation from Internet-based and Local Language Model Ensembles},
  author={Liu, Dan and others},
  booktitle={NAACL},
  year={2025},
  url={https://aclanthology.org/2025.naacl-long.259/}
}

@inproceedings{bijectionlearning2025,
  title={Endless Jailbreaks with Bijection Learning},
  author={Brian R. Y. Huang and Max Li and Leonard Tang},
  booktitle={International Conference on Learning Representations (ICLR)},
  year={2025},
  url={https://proceedings.iclr.cc/paper_files/paper/2025/hash/b05c1fb3345743dea59f500ec5a0bba0-Abstract-Conference.html}
}

@inproceedings{papineni-etal-2002-bleu,
    title = "{B}leu: a Method for Automatic Evaluation of Machine Translation",
    author = "Papineni, Kishore  and
      Roukos, Salim  and
      Ward, Todd  and
      Zhu, Wei-Jing",
    editor = "Isabelle, Pierre  and
      Charniak, Eugene  and
      Lin, Dekang",
    booktitle = "Proceedings of the 40th Annual Meeting of the Association for Computational Linguistics",
    month = jul,
    year = "2002",
    address = "Philadelphia, Pennsylvania, USA",
    publisher = "Association for Computational Linguistics",
    url = "https://aclanthology.org/P02-1040/",
    doi = "10.3115/1073083.1073135",
    pages = "311--318"
}

@article{nllb2022,
  title={No Language Left Behind: Scaling Human-Centered Machine Translation},
  author={{NLLB Team} and Costa-juss{\`a}, Marta R. and Cross, James and others},
  journal={arXiv preprint arXiv:2207.04672},
  year={2022},
  doi={10.48550/arXiv.2207.04672},
  url={https://arxiv.org/abs/2207.04672}
}

@article{nllb-finetune,
  title={Scaling neural machine translation to 200 languages},
  author={{NLLB Team}},
  journal={Nature},
  volume={630},
  pages={841--846},
  year={2024},
  doi={10.1038/s41586-024-07335-x},
  url={https://www.nature.com/articles/s41586-024-07335-x}
}

@article{malki2025hoovered,
  title={Hoovered up as a data point: Exploring Privacy Behaviours, Awareness, and Concerns Among {UK} Users of {LLM}-based Conversational Agents},
  author={Malki, Lisa Mekioussa and Polamarasetty, Akhil and Hatamian, Majid and Warner, Mark and Costanza, Enrico},
  journal={Proceedings on Privacy Enhancing Technologies},
  volume={2025},
  number={4},
  pages={838--860},
  year={2025},
  doi={10.56553/popets-2025-0160},
  url={https://doi.org/10.56553/popets-2025-0160}
}

@inproceedings{cox2025ephemerality,
  title={The Impact of a Chatbot's Ephemerality-Framing on Self-Disclosure Perceptions},
  author={Cox, Samuel Rhys and Jacobsen, Rune M{\o}berg and van Berkel, Niels},
  booktitle={CUI '25: Proceedings of the 2025 ACM Conference on Conversational User Interfaces},
  year={2025},
  doi={10.1145/3719160.3736617},
  publisher={Association for Computing Machinery},
  address={United States}
}

@misc{openai_data_controls_2025,
  title={Data controls in the OpenAI platform},
  author={{OpenAI}},
  year={2025},
  url={https://platform.openai.com/docs/guides/your-data},
  note={Accessed 2026-01-29}
}

@misc{openai_api_data_sharing_2025,
  title={Sharing feedback, evaluation and fine-tuning data, and API inputs and outputs with OpenAI},
  author={{OpenAI}},
  year={2025},
  url={https://help.openai.com/en/articles/9883556-sharing-model-feedback-through-the-api/},
  note={Accessed 2026-01-29}
}

@misc{anthropic_api_retention_2026,
  title={How long do you store my organization’s data?},
  author={{Anthropic}},
  year={2026},
  url={https://privacy.claude.com/en/articles/7996866-how-long-do-you-store-my-organization-s-data},
  note={Accessed 2026-01-29}
}

@misc{google_vertex_data_governance_2024,
  title={Vertex AI and zero data retention},
  author={{Google Cloud}},
  year={2024},
  url={https://cloud.google.com/vertex-ai/generative-ai/docs/data-governance},
  note={Accessed 2026-01-29}
}
